\documentclass[11pt,a4paper]{article}
\pdfoutput=1
\usepackage{jheppub}
\usepackage{bm}
\usepackage{graphicx,psfrag}
\usepackage{mathtools}
\usepackage{mathrsfs}
\usepackage{subfigure}
\usepackage[svgnames]{xcolor}
\usepackage{bm}
\usepackage{slashed}
\usepackage{enumerate}
\usepackage{xspace}

\newcommand{\kt}{\bm{k_{\perp}}}

\newcommand{\kp}{k_{\perp}}
\newcommand{\kpj}{k_{\perp\, j}}
\newcommand{\Ph}{P_{h\perp}}

\newcommand{\pp}{p_{\perp}}
\newcommand{\btb}{\bm{b}_T}
\newcommand{\btu}{b_T}
\newcommand{\bt}{{\bm{b}_{T}}}
\newcommand{\Sp}{S_{||}}
\newcommand{\Bt}{B_{T}}

\newcommand{\qt}{\bm{P}_{h\perp}}
\newcommand{\qtu}{{P}_{h\perp}}

\newcommand{\qti}{P_{h\perp i}}

\newcommand{\qtpj}{P^{[+]}_{h\perp {j}}}
\newcommand{\qtmj}{P^{[-]}_{h\perp {j}}}
\newcommand{\qtpmj}{P^{[\pm]}_{h\perp {j}}}
\newcommand{\ktpj}{k^{[+]}_{\perp {j}}}
\newcommand{\ktmj}{k^{[-]}_{\perp {j}}}

\newcommand{\bp}{b_{T}}
\newcommand{\pt}{\bm{p_{\perp}}}

\newcommand{\kx}{k_{x}}

\newcommand{\ky}{k_{y}}

\newcommand{\kz}{k_{z}}

\newcommand{\Phperp}{\bm{P}_{h\perp}}
\newcommand{\Php}{|\bm{P}_{h\perp}|}
\newcommand{\Phperpz}{\bm{P}_{h\perp}}

\newcommand{\TMDPs}{TMD PDFs\xspace}

\newcommand{\FTStrufu}{\mathcal{F}}
\newcommand{\tS}{\tilde{S}}
\newcommand{\tsigma}{\tilde{\sigma}}

\newcommand{\xbj}{x}                   

\newcommand{\dphi}{d\Phi}
\newcommand{\slim}{\mskip 1.5mu}              

\newcommand{\la}{\langle}
\newcommand{\ra}{\rangle}
\newcommand{\lf}{\left}
\newcommand{\rg}{\right}

\newcommand{\open}{{<\kern -0.3 em{\scriptscriptstyle )}}}

\usepackage{ulem}
\usepackage{enumerate}
\usepackage{xspace}

\newcommand{\Poi}{\mathrm{Poi}}
\newcommand{\Multi}{\mathrm{Multi}}

\newcommand{\be}{\begin{equation}}
\newcommand{\ee}{\end{equation}}
\newcommand{\bea}{\begin{eqnarray}}
\newcommand{\eea}{\end{eqnarray}}
\newcommand{\nn}{\nonumber\\}

\title{Studies of Transverse Momentum Dependent Parton Distributions
and Bessel Weighting}

\author[a,b]{M. Aghasyan,}
\author[c]{H. Avakian,}
\author[a]{E. De Sanctis,}
\author[d]{L. Gamberg,}
\author[a]{M. Mirazita,}
\author[e]{B. Musch,}
\author[c]{A. Prokudin,}
\author[a,c]{P. Rossi}

\affiliation[a]{INFN, Laboratori Nazionali di Frascati, 00044 Frascati, Italy} 
\affiliation[b]{Instituto Tecnol\'ogico da Aeron\'autica/DCTA 12228-900, S\~ao Jos\'e dos Campos, SP, Brazil}
\affiliation[c]{Jefferson Lab, 12000 Jefferson Avenue,  Newport News, Virginia 23606, USA}
\affiliation[d]{Department of Physics,    Penn State University-Berks,  Reading, PA 19610, USA}

\affiliation[e]{Institut f\"ur Theoretische Physik, Universit\"at Regensburg, 
 93040 Regensburg, Germany}

\emailAdd{mher@jlab.org}
\emailAdd{avakian@jlab.org}
\emailAdd{Enzo.DeSanctis@lnf.infn.it}
\emailAdd{lpg10@psu.edu}
\emailAdd{marco.mirazita@lnf.infn.it}
\emailAdd{bmusch@ph.tum.de}
\emailAdd{prokudin@jlab.org}
\emailAdd{rossi@jlab.org}

\arxivnumber{JLAB-THY-14-1945}

\abstract{In this paper 
we present a new technique for analysis of 
 transverse momentum dependent  parton distribution functions, 
based on the Bessel weighting formalism. The procedure is applied to studies of  the double longitudinal 
spin asymmetry in semi-inclusive deep inelastic 
scattering  using a new dedicated Monte Carlo generator  which includes quark intrinsic transverse momentum  within the generalized parton model. 
Using a fully differential cross section for the process, 
the effect of four momentum conservation is analyzed 
using various input models for transverse momentum 
 distributions and fragmentation functions.
We observe a few percent systematic offset of the Bessel-weighted asymmetry obtained from Monte Carlo extraction compared to input model calculations, which is due to the limitations imposed by the energy and momentum conservation at the given energy/$Q^2$. We find that the Bessel weighting technique provides a powerful and reliable tool to study the Fourier transform of TMDs with controlled systematics due to experimental acceptances and resolutions with different TMD 
model inputs.}

\keywords{SIDIS, parton intrinsic transverse momentum, azimuthal moments}

\begin{document}  

\maketitle

\section{Introduction}
\label{se-intro}
The study of the  spin structure of protons and neutrons is one of the central issues in hadron physics, with many dedicated experiments, recent 
(HERMES at DESY, CLAS and Hall-A at JLAB), 
running (COMPASS at CERN, STAR and PHENIX at RHIC), approved (JLab $12~{\rm GeV}$ upgrade \cite{Dudek:2012vr}, COMPASS-II~\cite{Gautheron:2010wva}) or planned (Electron Ion  Collider \cite{Gao:2010av,Boer:2011fh,Accardi:2012qut}). 
The Transverse Momentum Dependent (TMD) parton distribution functions  
and fragmentation functions play a crucial role in gathering and 
interpreting information of a true ``3-dimensional'' imaging of the nucleon. 
These Transverse Momentum Dependent distribution and fragmentation functions (collectively here called ``TMDs'') 
can be accessed in several types of processes, one of the most important 
 is single particle hadron production in  Semi-Inclusive Deep Inelastic Scattering (SIDIS) of 
leptons on nucleons.
 A significant amount of data on spin-azimuthal distributions of hadrons in SIDIS, providing access to TMDs has been accumulated in recent years by 
several collaborations, including HERMES, COMPASS and Halls A,B and C at 
JLab~\cite{Airapetian:2004tw,Alexakhin:2005iw,Ageev:2006da,Airapetian:2010ds,Alekseev:2010rw,Alekseev:2010dm,Adolph:2012nw,Mkrtchyan:2007sr,Avakian:2010ae,Qian:2011py}.
At least an order of magnitude more data is expected in coming years of running of JLab 12~\cite{Dudek:2012vr}. 

A rigorous basis for studies of TMDs in SIDIS is provided by
TMD factorization in QCD, which has been established in 
Refs.~\cite{Collins:1981uk,Collins:1981uw,Collins:1984kg,Ji:2004wu,Collins:2004nx, Bacchetta:2008xw,Aybat:2011zv,Collins:2011zzd} for 
leading twist  
single hadron production with transverse momentum 
of the produced hadron 
being much smaller than the hard scattering scale, and the order of
$\Lambda_{QCD}$, that is
 $\Lambda_{QCD}^2 < P_{h\perp}^2 \ll Q^2$. 
In this kinematic domain  the SIDIS cross section  can be expressed in terms of 
structure functions encoding the strong-interaction dynamics of
the hadronic sub-process $\gamma^* + p\to h + X$  
\cite{Kotzinian:1994dv,Mulders:1995dh,Levelt:1994np,Bacchetta:2006tn}, which are given by  convolutions of a hard scattering cross section and TMDs. 
However the extraction of TMDs as a function of the light-cone fraction $x$ 
and transverse momentum $\kp$ from single and double spin azimuthal asymmetries is hindered by the fact that observables are complicated convolutions in momentum space making the flavor decomposition of the 
underlying TMDs a model dependent procedure.

Based on TMD factorization theorems, experimentally measured cross sections are expressed as 
convolutions of TMDs where  $\kp$ dependence is integrated over and related to the measured value of   $P_{h\perp}$. A reliable method to directly access the 
$\kp$ dependence of TMDs is very desirable.  However,  various assumptions involved in modern extractions of TMDs from available data 
rely on  conjectures  of the  transverse momentum dependence of distribution and  fragmentation functions ~\cite{deFlorian:2007aj,Anselmino:2005nn,
Amrath:2005gv,Bacchetta:2007wc,PhysRevD.83.074003,Hirai:2007cx,
Anselmino:2008sga,Matevosyan:2012ga,Matevosyan:2011vj,Casey:2012hg,Signori:2013mda}
 making  estimates 
of systematic errors due to those assumptions
extremely challenging. 

In a  paper by Boer, Gamberg, Musch,  and Prokudin~\cite{Boer:2011xd}, a new technique has been proposed called Bessel weighting,
 which relies on a model-independent deconvolution of structure functions in terms of Fourier  transforms  of TMDs from observed azimuthal moments in SIDIS with polarized and unpolarized targets.  In this paper,  we apply the Bessel weighting procedure 
to  present an extraction of Fourier transforms of TMDs 
from a   Monte Carlo event generator.
As an application of this procedure  we consider 
the ratio of helicity $g_{1L}$,   and 
unpolarized $f_1$ TMDs from  the double longitudinally polarization asymmetry.

This paper is organized as follows: 
We begin our discussion in Section~\ref{sec:bessel-weighting} with 
a brief review of the formalism of the SIDIS cross section and its
representation in both momentum and Fourier conjugate $\bp$ space.
 The latter representation  lends itself
to a discussion of the Bessel weighting formalism~\cite{Boer:2011xd}.
 We review its merits in studying the transverse structure of the nucleon
and  present a description of the experimental procedure to study TMDs using 
Bessel weighting which provides  a new tool to study nucleon structure.
 In Section~\ref{sec:mc} we introduce a fully differential  Monte Carlo  generator which  has been developed to test the procedure for extraction of TMDs from SIDIS.    
As a test of the quality of  our constructed 
 Monte Carlo, in  Section~\ref{Cahn} we present a study of the 
 Cahn effect~\cite{Cahn:1978se,Cahn:1989yf} contribution to the
average  $\la\cos\phi\ra$ moment in SIDIS. 
  In Section~\ref{sec:extract}
we present the extraction of the double spin asymmetry $A_{LL}(\bp)$, defined as the ratio of the difference and the sum of electro-production cross sections for anti-parallel and parallel configurations of lepton and nucleon spins
 using the Bessel weighting procedure.  The effects of different model inputs and experimental resolutions and acceptances on extracted TMDs  are investigated.
   Finally in Section~\ref{sec:conc} we draw some conclusions of the
present analysis and outline steps for future work.

\section{Extraction of TMDs using Bessel Weighting}
\label{sec:bessel-weighting}
\subsection{The Cross Section for Semi-inclusive Deep Inelastic Scattering}
 The SIDIS cross section can be
 expressed in a model independent way in terms of a  set of 18 structure
functions~\cite{Gourdin:1973qx,Kotzinian:1994dv,Mulders:1995dh,Boer:1997nt,Diehl:2005pc,Bacchetta:2006tn},
\begin{align}
\lefteqn{
\frac{d\sigma}{d\xbj \, dy\, d\psi \,dz\, d\phi_h\, d| \Phperp|^2}
 = \frac{\alpha^2}{\xbj y\slim Q^2}\,
\frac{y^2}{2\,(1-\varepsilon)}\,  \biggl( 1+\frac{\gamma^2}{2\xbj} \biggr)
\Biggl\{F_{UU ,T} + \varepsilon\slim F_{UU ,L}}
\nonumber \\  &  \quad \quad
+\quad \sqrt{2\,\varepsilon (1+\varepsilon)}\,\cos\phi_h\, F_{UU}^{\cos\phi_h}
+ \varepsilon \cos(2\phi_h)\, 
F_{UU}^{\cos 2\phi_h}
\nn  & \quad \quad
+ \lambda_e\, \sqrt{2\,\varepsilon (1-\varepsilon)}\, 
           \sin\phi_h\, 
F_{LU}^{\sin\phi_h}
\phantom{\Bigg[ \Bigg] }
\nonumber \\  & \quad \quad
+ S_\parallel\, \Bigg[ 
 \sqrt{2\, \varepsilon (1+\varepsilon)}\,
  \sin\phi_h\, 
F_{UL}^{\sin\phi_h}
+  \varepsilon \sin(2\phi_h)\, 
F_{UL}^{\sin 2\phi_h}
\Bigg]
\nonumber \\  & \quad \quad
+ S_\parallel \lambda_e\, \Bigg[ \,
  \sqrt{1-\varepsilon^2}\; 
F_{LL}
+\sqrt{2\,\varepsilon (1-\varepsilon)}\,
  \cos\phi_h\, 
F_{LL}^{\cos \phi_h}
\Bigg]
\nonumber \\  & \quad \quad
+ |\bm{S}_\perp|\, \Bigg[
  \sin(\phi_h-\phi_S)\,
\Bigl(F_{UT ,T}^{\sin\lf(\phi_h -\phi_S\rg)}
+ \varepsilon\, F_{UT ,L}^{\sin\lf(\phi_h -\phi_S\rg)}\Bigr)
\nonumber \\  & \quad  \quad \quad
+ \varepsilon\, \sin(\phi_h+\phi_S)\, 
F_{UT}^{\sin\lf(\phi_h +\phi_S\rg)}
+ \varepsilon\, \sin(3\phi_h-\phi_S)\,
F_{UT}^{\sin\lf(3\phi_h -\phi_S\rg)}
\phantom{\Bigg[ \Bigg] }
\nonumber \\  & \quad \quad \quad
+ \sqrt{2\,\varepsilon (1+\varepsilon)}\, 
  \sin\phi_S\, 
F_{UT}^{\sin \phi_S }
+ \sqrt{2\,\varepsilon (1+\varepsilon)}\, 
  \sin(2\phi_h-\phi_S)\,  
F_{UT}^{\sin\lf(2\phi_h -\phi_S\rg)}
\Bigg]
\nonumber \\  & \quad \quad 
+ |\bm{S}_\perp| \lambda_e\, \Bigg[
  \sqrt{1-\varepsilon^2}\, \cos(\phi_h-\phi_S)\, 
F_{LT}^{\cos(\phi_h -\phi_S)}
+\sqrt{2\,\varepsilon (1-\varepsilon)}\, 
  \cos\phi_S\, 
F_{LT}^{\cos \phi_S}
\nonumber \\  & \quad \quad \quad
+\sqrt{2\,\varepsilon (1-\varepsilon)}\, 
  \cos(2\phi_h-\phi_S)\,  
F_{LT}^{\cos(2\phi_h - \phi_S)}
\Bigg] \Biggr\},
\label{e:crossmaster}
\end{align}
where the first two subscripts of the structure functions $F_{XY}$ 
indicate the polarization of the beam and target, and in certain
cases, a third sub-script in $F_{XY,Z}$ indicates  the polarization of the 
virtual photon. The structure functions depend 
on the  
the scaling variables $\xbj$, $z$, the four momentum $Q^2=-q^2$,
where  $q=l-l^{'}$ is the momentum of the virtual photon, 
and  $l$ and $l^{'}$  
are the 4-momenta  of the incoming and outgoing leptons, respectively. 
$P_{h\perp}$ is the transverse momentum component of the produced hadron 
 with respect to the virtual photon direction.  
 
The scaling variables 
have the standard definitions, 
$\xbj = Q^2/{2(P\cdot q)}$, $y={(P\cdot q)/(P\cdot l)}$, and $z={(P\cdot P_h)/(P\cdot q)}$. Further, in Eq.~(\ref{e:crossmaster}) 
 $\alpha$ is the fine structure constant; the angle $\psi$ is the azimuthal angle of $\ell'$
around the lepton beam axis with respect to an arbitrary fixed
direction~\cite{Diehl:2005pc},   
 and   $\phi_h$ is the 
azimuthal angle between the scattering plane formed by the initial
 and final momenta of the electron and the production plane formed 
by the transverse momentum of the  observed hadron
and the virtual photon, whereas 
$\phi_S$ is the azimuthal angle of the transverse 
spin in the scattering plane~\cite{Bacchetta:2004jz}.
Finally, $\varepsilon$ is the  ratio of longitudinal and
transverse photon fluxes~\cite{Bacchetta:2006tn}.

At tree-level,  
in a parton model factorization framework~\cite{Mulders:1995dh,Boer:1997nt,Bacchetta:2006tn},  
the various structure functions in the cross section are written as convolutions of the  TMDs  
which relate transverse momenta
of the active partons and produced hadron.
 For our purposes, the unpolarized and double longitudinal polarized
structure functions are
\bea
F_{UU,T} &=& 
x\, \sum_a e_a^2 \int d^2 \bm{\pt}\,  d^2 \bm{\kt}^{}
\, \delta^{(2)}\bigl(z\bm{\kt}+ \bm{\pt}  - \bm{P}_{h \perp} \bigr)
\,
f_1^a(x,\kt^2)\,D_1^a(z,\pt^2)\, \label{conv0} ,\\ 
F_{LL}
&=& x\,
\sum_a e_a^2 \int d^2 \bm{\pt}\,  d^2 \bm{\kt}^{}
\, \delta^{(2)}\bigl(z\bm{\kt}+ \bm{\pt}  - \bm{P}_{h \perp} \bigr)
\,
g_{1L}^a(x,\kt^2)\,D_1^a(z,\pt^2) ,
\label{conv}
\eea
where $\kt$ is the intrinsic transverse momentum of the struck quark, and
 $\pt$ is the transverse momentum  of the final state hadron
relative to the fragmenting quark $k^\prime$ (see Fig.~\ref{kinem}).   
$f_1^a(x,\kt^2)$, $g_{1L}^2$ and $D^a(z,\pt^2)$ represent TMD PDFs and fragmentation functions  respectively of flavor $a$, 
$e_a$ is the fractional charge of the struck quark or anti-quark and
the summation runs over quarks and anti-quark flavors $a$.

Measurements of the transverse momentum $P_{h\perp}$ of 
final state hadrons in SIDIS with polarized leptons and nucleons provide access to transverse momentum dependence of  TMDs.
Recent measurements of multiplicities and double spin asymmetries as a function of the final transverse momentum of pions in SIDIS at COMPASS \cite{Adolph:2013stb}, HERMES \cite{Airapetian:2012ki}, and JLab \cite{Mkrtchyan:2007sr,Avakian:2010ae,Qian:2011py} suggest that transverse momentum distributions depend on the polarization of quarks and possibly  also on their flavor~\cite{Signori:2013mda}  (see also discussion in Ref.~\cite{Anselmino:2013lza}). Calculations of transverse momentum dependence of TMDs in different models \cite{Lu:2004au,Anselmino:2006yc,Pasquini:2008ax,Bourrely:2010ng} and on the lattice \cite{Hagler:2009mb,Musch:2010ka} also indicate that the dependence of  transverse momentum  distributions on the quark polarization and flavor maybe significant. 
Larger intrinsic transverse momenta of sea-quarks compared to valence 
quarks have been discussed in an effective model of the low energy dynamics 
resulting from chiral symmetry breaking in QCD~\cite{Schweitzer:2012hh}. 

As stated above, the various 
assumptions 
on  transverse momentum dependence of
distributions on spin and flavor of 
quarks however make phenomenological fits very challenging. To 
minimize these model assumptions, Kotzinian and Mulders \cite{Kotzinian:1997wt} suggested using so called $P_{h\perp}$-weighted asymmetries, where the unknown 
$\kp$-dependencies of TMDs are 
integrated out,  thus 
providing access to  moments of TMDs. However, the  
$P_{h\perp}$-weighted asymmetries introduce a significant challenge to both theory and experiment. For example, the weighting with $P_{h\perp}$ 
emphasizes the kinematical region with higher $\Ph$, 
 where the statistics are  poor and systematics from detector acceptances are
 difficult to control and at the same time theoretical description in terms of TMDs breaks down. 

The method of Bessel weighting~\cite{Boer:2011xd} addresses these experimental and theoretical issues.  First, 
Bessel weighted asymmetries are given 
in terms of simple products of  Fourier transformed
TMDs  without imposing any model assumptions of the  their transverse momentum dependence.  Secondly,  Bessel weighting regularizes the ultraviolet divergences
resulting from unbound  momentum integration that arises from conventional
weighting. Further, in this paper we will demonstrate that they provide a new experimental
tool to study the TMD content to the SIDIS cross section that 
minimize the transverse momentum model dependencies inherent in conventional
extractions of TMDs. Also they  suppress the kinematical regions where
cross sections are small and statistics are poor~\cite{Boer:2011xd}.

We begin the discussion of Bessel weighting  by 
 re-expressing  the SIDIS cross section 
as a Bessel weighted integral in $\btu$ space~\cite{Boer:2011xd}:
\begin{align}
\lefteqn{\frac{d\sigma}{d\xbj \, dy\, d\psi \,dz_h\, d\phi_h\, d |\Phperp|^2} = } 
\nonumber \\   & \quad
\frac{\alpha^2}{\xbj y\slim Q^2}\,
\frac{y^2}{(1-\varepsilon)}\,  \biggl( 1+\frac{\gamma^2}{2\xbj} \biggr)\, \int \frac{d |\btb|}{(2\pi)} |\btb|\Biggl\{
J_{0} (|\btb| |\Phperpz|)\, \FTStrufu_{UU,T}  
+ 
\varepsilon\slim
J_{0} (|\btb| |\Phperpz|)\, \FTStrufu_{UU ,L}
\nonumber \\  & \quad
+ \quad \sqrt{2\,\varepsilon (1+\varepsilon)}\,\cos\phi_h\,
J_{1} (|\btb| |\Phperpz|)\,\FTStrufu_{UU}^{\cos\phi_h}
 + 
\varepsilon \cos(2\phi_h)\, 
J_{2} (|\btb| |\Phperpz|)\, \FTStrufu_{UU}^{\cos(2\phi_h)} 
\nonumber \\  & \quad
+ \lambda_e\, \sqrt{2\,\varepsilon (1-\varepsilon)}\, 
           \sin\phi_h\, 
J_{1} (|\btb| |\Phperpz|)\,\FTStrufu_{LU}^{\sin\phi_h}
\phantom{\Bigg[ \Bigg] }
\nonumber \\  & \quad 
+ \quad S_\parallel\, \Bigg[
\sqrt{2\, \varepsilon (1+\varepsilon)}\,
  \sin\phi_h\, 
J_{1} (|\btb| |\Phperpz|)\,\FTStrufu_{UL}^{\sin\phi_h}
+  \varepsilon \sin(2\phi_h)\, 
J_{2} (|\btb| |\Phperpz|)\,\FTStrufu_{UL}^{\sin 2\phi_h}
\Bigg]
\nonumber \\  &  \quad
+ S_\parallel \lambda_e\, \Bigg[ \,
  \sqrt{1-\varepsilon^2}\,
J_{0} (|\btb| |\Phperpz|)\, \FTStrufu_{LL}
+\sqrt{2\,\varepsilon (1-\varepsilon)}\,
  \cos\phi_h\, 
J_{1} (|\btb| |\Phperpz|)\, \FTStrufu_{LL}^{\cos\phi_h} 
\Bigg]
\nonumber \\ &  \quad
 +
|\bm{S}_\perp|\, \Bigg[
  \sin(\phi_h-\phi_S)\, J_{1} (|\btb| |\Phperpz|)\, \,
\Bigl(\FTStrufu_{UT,T}^{\sin(\phi_h-\phi_S)}  
+ \varepsilon\, \FTStrufu_{UT ,L}^{\sin\lf(\phi_h -\phi_S\rg)}\Bigr)
\nonumber \\ &  \quad \qquad
+ 
\; \varepsilon\, \sin(\phi_h+\phi_S)\, 
J_{1} (|\btb| |\Phperpz|)\, \FTStrufu_{UT}^{\sin(\phi_h+\phi_S)}
\nonumber \\  & \quad \qquad 
+
\; \varepsilon\, \sin(3\phi_h-\phi_S)\,
J_{3} (|\btb| |\Phperpz|)\, \FTStrufu_{UT}^{\sin\lf(3\phi_h -\phi_S\rg)}
\phantom{\Bigg[ \Bigg] }
\nonumber \\  & \quad \qquad 
+
\; \sqrt{2\,\varepsilon (1+\varepsilon)}\, 
  \sin\phi_S\, 
J_{1} (|\btb| |\Phperpz|)\, \FTStrufu_{UT}^{\sin \phi_S }
\nonumber \\  & \quad \qquad
 +
\;  \sqrt{2\,\varepsilon (1+\varepsilon)}\, 
  \sin(2\phi_h-\phi_S)\,  
J_{2} (|\btb| |\Phperpz|)\, \FTStrufu_{UT}^{\sin\lf(2\phi_h -\phi_S\rg)}
\Bigg]
\nonumber \\  &  \quad 
+
|\bm{S}_\perp| \lambda_e\, \Bigg[
  \sqrt{1-\varepsilon^2}\, \cos(\phi_h-\phi_S)\, 
J_{1} (|\btb| |\Phperpz|)\, \FTStrufu_{LT}^{\cos(\phi_h -\phi_S)}
\nonumber \\  &  \quad \qquad 
+
\; \sqrt{2\,\varepsilon (1-\varepsilon)}\, 
  \cos\phi_S\, 
J_{0} (|\btb| |\Phperpz|)\, \FTStrufu_{LT}^{\cos \phi_S}
\nonumber \\  &  \quad \qquad  
+
\; \sqrt{2\,\varepsilon (1-\varepsilon)}\, 
  \cos(2\phi_h-\phi_S)\,  
J_{2} (|\btb| |\Phperpz|)\, \FTStrufu_{LT}^{\cos(2\phi_h - \phi_S)}
\Bigg] \Biggr\}
\label{eq:CS_bspace}
\end{align}
where  in the parton model framework 
the structure functions $\FTStrufu_{XY,Z}$ are now given as
 simple {\it products} of Fourier Transforms of TMDs.   
Here  we consider   the unpolarized and  double longitudinal
structure functions,
\begin{align}
  \FTStrufu _{UU,T} &= x\, \sum_a e_a^2 \tilde f_1^a(x, z^2\bt^2)\tilde D_1^a(z, \bt^2)\, , \\
\FTStrufu_{LL}  &=  x\, \sum_a e_a^2 \tilde g_{1L}^a(x, z^2\bt^2) \tilde D_1^a(z, \bt^2)\, , 
\label{eq:strucutt}
\end{align}
 where the Fourier transform of the TMDs  are defined as
\begin{align}
\tilde f(x, \bt^2) & =  \int  d^2 \kt \, e^{i \bt\cdot \kt }\; f(x, \kt^2)  
 = 2\pi\hskip -0.15cm \int   d \kt\kt J_0(|\bt||\kt|)\; f(x, \kt^2)\, 
\label{eq-bttransf}\, ,
\end{align}
\begin{align}
\tilde D(z, \bt^2)  & =  \int d^2 \pt \,  e^{i  \bt\cdot \pt }\; D(z, \pt^2) 
 = 2\pi\hskip -0.15cm \int d \pt\pt J_0(|\bt||\pt|)\; D(x, \pt^2)\, .
\end{align}

\subsection{Bessel Weighting of Experimental Observables}

In this sub-section we introduce Bessel weighting of experimental observables,
cross sections and asymmetries,  based on the  $\btu$ representation of the SIDIS cross section, Eq.~(\ref{eq:CS_bspace}).
In a partonic framework, ``Bessel weighted  experimental observables'' 
 are quantities which can be presented as simple products of Fourier transforms  of distribution and fragmentation functions, allowing the application of standard flavor decomposition procedures.    Here we will apply this technique to the double longitudinal spin
asymmetry.  From Eq.~(\ref{eq:CS_bspace}) 
 one can project out the unpolarized and double longitudinally polarized structure functions $\FTStrufu_{LL}$, and   $\FTStrufu _{UU,T}$,  
by integrating  with the zeroth order Bessel function
$J_{0} (|\btb| |\Phperpz|)$ over the transverse momentum of the produced
hadron $\Phperp$.  We arrive at an expression for the longitudinally polarized cross section $\tilde{\sigma}^{\pm}(\bp)$
in $\bt$-space
\bea
 \tilde \sigma^{\pm}(b_T) =2\pi \int \frac{ d \sigma^{\pm} }{d\Phi}J_{0}(|\btb| |\Phperpz|
) \Ph \, d\Ph ,
\label{eq:project}
\eea 
where   $d\Phi\equiv dx\, dy\, d\psi\,  dz\,  d\Ph P_{h\perp}$
 represents shorthand notation for the phase space differential  and 
$|\btb|\equiv\btu$, and $|\Phperp|\equiv P_{h\perp}$, 
$d \sigma^{\pm}/d\Phi$ is the differential cross section where
$\pm$ labels the double longitudinal spin combinations $\Sp\lambda_e=\pm 1$. 
 Note that in our definition $\btu$ is the Fourier conjugate variable to $P_{h\perp}$
~\cite{Boer:2011xd}.

Now we form  the double longitudinal spin asymmetry  
\bea
A^{J_{0}(b_T\Ph)}_{LL}(b_T) = \frac{ \tilde \sigma^+(b_T) - \tilde \sigma^-(b_T)}{\tilde \sigma^+(b_T) + \tilde \sigma^-(b_T) }\equiv\frac{\tilde \sigma_{LL}(b_T)}{\tilde \sigma_{UU}(b_T)}=\sqrt{1-\varepsilon^2} \frac{\sum_{a} e_a^2 \tilde g^{a}_{1L}(x,z^2\btu^2) \tilde D^{a}_{1}(z,\btu^2)}{\sum_{a} e_a^2\tilde f^{a}_1(x,z^2\btu^2) \tilde D^{a}_{1}(z,\btu^2)}\, .   \nn
\label{tildall}
\eea 
The experimental procedure to study the 
structure functions in $\btu$-space  
amounts to discretizing the momentum phase space in Eq.~(\ref{eq:project}) and constructing the sums and differences of these discretized cross sections.
The technical details of  this procedure  given in Appendix~\ref{Projectbin} and \ref{errors}.  
Using these results, the double longitudinal spin asymmetry,   
Eq.~(\ref{tildall}) 
results in an expression of sums and differences 
 of Bessel functions for a given set of  
experimental events. 
The resulting expression for the spin asymmetry is
\bea
A^{J_{0}(b_T\Ph)}_{LL}(b_T) &=& 
\frac{\displaystyle\sum_{{j}}^{N^{+}}\,  J_0(\btu \qtpj) 
-
\sum_{{j}}^{N^{-}}\,  J_0(\btu \qtmj)}
{\displaystyle\sum_{{j}}^{N^{+}}\,  J_0(\btu \qtpj)
+\sum_{{j}}^{N^{-}}\,  J_0(\btu \qtmj)}\, ,
\label{spma1}
\eea
 where $j$ indicates a sum on $\pm$-helicity  events\footnote{Note,
the $+$ helicity and $-$ helicity events are in two different, 
independent data sets of transverse momenta.},  and  
where  $N^{\pm}$ is the number of events with positive/negative products of
lepton and nucleon helicities.

The cross sections $\tilde \sigma^\pm(b_T)$ can be extracted for any given $\btu$ using sums over the same set of data. These cross sections contain the same information as the cross sections,  $d\sigma/d\Phi$ in Eq.~(\ref{eq:project}) differential with respect to the outgoing hadron momentum. 
The momentum dependent and the $\btu$-dependent representations of the cross section are related by a 2-D Fourier-transform in cylinder coordinates. 
 Eq.~(\ref{spma1}) and its generalization to other 
spin and azimuthal asymmetries provides  another lever arm to study 
the partonic content of hadrons through the Bessel weighing procedure in Fourier $\btu$ space (See also \cite{Konychev:2005iy,Aidala:2014hva}).

In order to test the Bessel weighting of experimental observables for 
the double longitudinal spin asymmetry 
we will use   a Monte Carlo  generator which  has been developed 
for  the extraction of TMDs from SIDIS.  In the next Section we describe this  new dedicated Monte Carlo generator 
which includes quark intrinsic transverse momentum within the generalized
parton model.

\section{Fully Differential Monte Carlo for SIDIS}
\label{sec:mc}

\subsection{The Monte Carlo and the Generalized Parton Model}
A Monte Carlo generator is a  crucial component in testing  experimental procedures such as those described in Eq.~(\ref{spma1}).
In order  to check the Bessel weighting technique 
we need a Monte Carlo that generates events in phase space 
with different TMD model inputs. It should also
include  explicit dependence on
intrinsic parton transverse momentum 
 $\kp$ and $\pp$. We  
 reconstruct   weighted asymmetries according to Eq.~(\ref{spma1}),  
and in turn compare the generated events in momentum space which are 
then  Fourier transformed.  
  In keeping with the parton model picture however, a   cross-section based on structure functions from Eqns.~(\ref{conv0}) and (\ref{conv})    cannot be used for these purposes, since the simple parton model factorization
 would allow the MC generator to produce events that violate four-momentum conservation and thus are unphysical.

 Therefore, the Monte Carlo generator we use has been  
developed to study partonic intrinsic motion using 
 the framework of the so-called generalized parton model described in detail in Ref.~\cite{Anselmino:2005nn}.  While including target mass corrections, more importantly for our study, it generates only events allowed by
the available  physical phase space.
  
\begin{figure}[t]
\centerline{\includegraphics[height=2.0in]{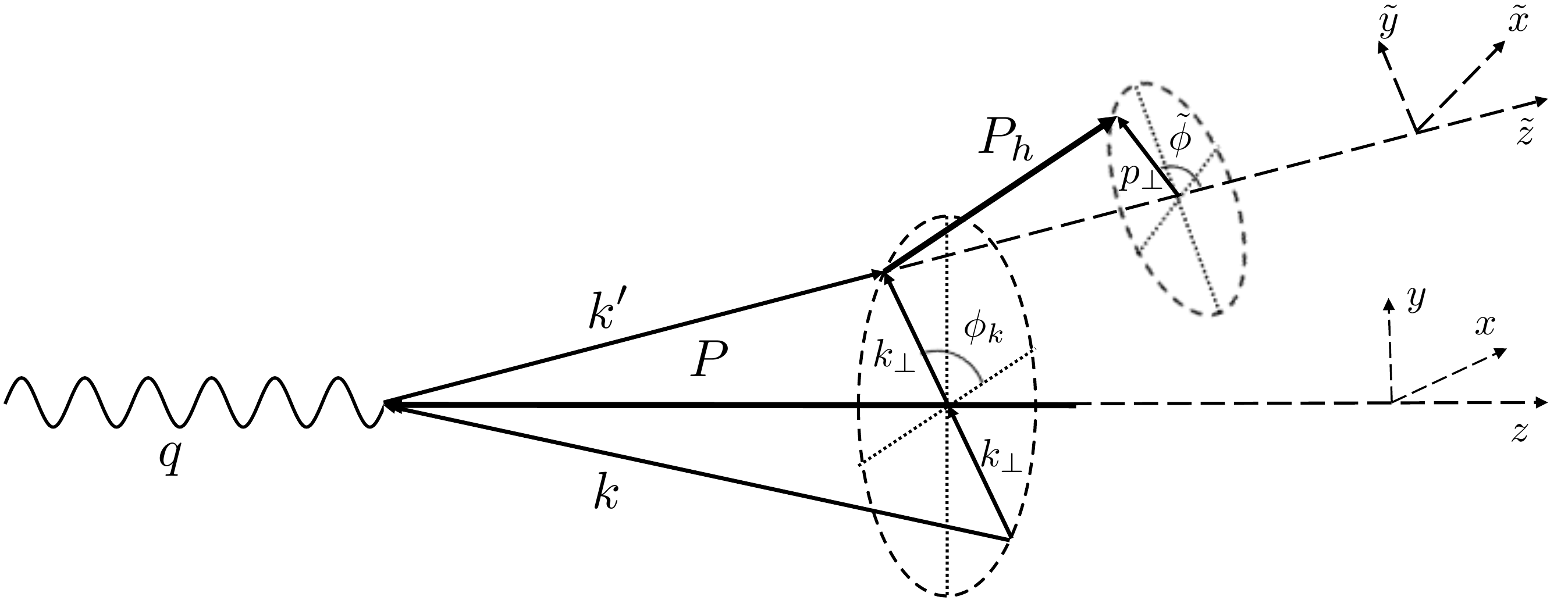}}
\caption{\scriptsize Kinematics of the process. $q$ is the virtual photon, $k$ and $k'$ are the initial and struck quarks, $k_{\perp}$ is the quark transverse component. $P_h$ is the final hadron with a $p_{\perp}$ component, transverse with respect to the fragmenting quark $k'$ direction. }
\label{kinem}
\end{figure}

In order to establish the proper kinematics of the  phase space
for the Monte Carlo   consider the SIDIS process
\be
{\ell}(l) + N(P)\rightarrow \ell(l') + h(P_{h}) + X, 
\ee
where $\ell$ is the incident lepton, $N$ is the target nucleon, and $h$ represents the observed hadron, and the four-momenta are given in parenthesis. Following the Trento conventions \cite{Bacchetta:2004jz}, the spatial component of
the virtual photon momentum $q$ is along the positive $z$  direction and the proton momentum $P$ is in the opposite direction, as depicted in  Fig.~\ref{kinem}. 
In the parton model, the virtual photon scatters off an on-shell quark where  the initial quark momentum $k$,  and scattered quark momentum $k^\prime$, have the same intrinsic transverse momentum component $\bm{\kp}$ with respect to the $z$ axis,  and where the  initial quark has 
the   fraction $x$ of the proton momentum.   The produced hadron momentum, $P_h$ has the fraction $z$ of scattered quark momentum  $k^\prime$  in the $(\tilde x,\tilde y, \tilde z)$ frame and  $\pt$ is the transverse momentum component 
 with respect to the scattered quark $k^\prime$. 

A great deal of phenomenological effort 
 has been devoted to using the 
generalized parton model  (see for example~\cite{Anselmino:2005nn,Anselmino:2008sga,Schweitzer:2010tt}), incorporating  intrinsic quark
transverse momentum, to account for experimentally 
observed spin   and azimuthal asymmetries as a function of the produced 
 hadron's   transverse momentum  $P_{h\perp}$  
in SIDIS processes.  In order to take into account  non-trivial   kinematic effects   that are  neglected  from the standard 
parton model approximations~\cite{Mulders:1995dh,Bacchetta:2006tn},  
such as discarding  small momenta in the struck and fragmenting
quarks,   and discarding transverse momentum kinematic corrections due 
to hard scattering we develop a Monte Carlo  based on 
the fully differential SIDIS cross section
~\cite{Anselmino:2005nn} which is given by,
\bea
\frac{d \sigma}{ dx dy dz  d \pp^2  d \kp^2 d \phi_{l^\prime} d \phi_k d \tilde \phi }& = &\frac{1}{2}K(x,y)J(x,Q^2,\kt^2)  \nn
&& \hspace{ -4.5cm} \times\,  x \sum_a e_a^2 \left[  f_{a}(x_{LC},\kt^2)D_{1,a}(z_{LC},\pt^2)  + \lambda \sqrt{1-\varepsilon^2}g_{1L,a}(x_{LC},\kt^2)D_{1,a}(z_{LC},\pt^2) \right] \, ,\nn
\label{FDALL}
\eea
 where the summation runs over quarks flavors and,
$\lambda$ is the product of target polarization and beam helicity 
($\lambda=\pm1$), $\phi_{l^\prime}$ is the scattered lepton azimuthal angle~\footnote{Integration 
over $\phi_{l^\prime}$ gives $2\pi$, since everything is symmetric along beam direction, although we need to keep it for further analysis, when one  reconstructs generated events in the real experimental setup.}, and 
\be
\begin{aligned}
K(x,y) = \frac{\alpha^2}{x y Q^2} \frac{y^2}{2(1- \varepsilon)}  \left( 1+\frac{\gamma^2}{2x} \right)\,, \quad
\varepsilon = \frac{1-y-\frac{1}{4}\gamma^2y^2}{1-y+\frac{1}{2}y^2+\frac{1}{4}\gamma^2 y^2} \,,
\label{Kfact}
\end{aligned} 
\ee
and the Jacobian $J$ is given by
\be
J(x,Q^2,\kt^2)=\frac{x}{x_{LC}} \left( 1+\frac{x^2}{x_{LC}^2}\frac{\kt^2}{Q^2}  \right) ^{-1}\, .
\ee

Here the cross section is  ``fully differential'' 
in the transverse momentum of the target and fragmenting quark. 
This form of the cross section will allow us to implement 
the physical energy and momentum 
phase space constraints in the Monte Carlo generator.
  In order to calculate the cross-section in terms of observed momenta (only linear combinations of $\kt$ and $\pt$ can be measured experimentally)
 we need to integrate Eq.~(\refeq{FDALL}) in $d^2\kt d^2\pt$ taking into account kinematical relations consistent with the  observed final hadron momentum
 $\Ph$.

We elaborate further on the kinematics for the Monte Carlo generator. 
 In above equations $x$ is the Bjorken variable, while $x_{LC}=k^-/P^-$ is the light-cone (LC) fraction of the proton momentum carried by the quark $k$ \cite{Anselmino:2005nn}. The quark four momentum  is given by,
\begin{equation}
k_0=x_{LC} P^{\prime} +\frac{k_{\perp}^2}{4x_{LC}P^{\prime}}, 
\end{equation}
\begin{equation}
\kx= k_{\perp} \cos(\phi_k), \quad 
\ky = k_{\perp} \sin(\phi_k), \quad
\kz= -x_{LC} P^{\prime} +\frac{k_{\perp}^2}{4x_{LC}P^{\prime}}, 
\end{equation}
where $k_0$ is the quark energy, and $k_{\left\{x,y,z\right\}}$ are the
 $x$, $y$ and $z$ components
of the quark momentum  in the center of mass (CM) frame of virtual photon and proton,   and
 $P^{\prime}\equiv 0.5(E_{p}+|P_{pz}|)$, where the proton energy in the CM is $E_{p} = \sqrt{P^2_{pz}+M^2}$.
Taking into account the nucleon mass, and 
the on-shell condition for the intial quark, the following expressions
for $x_{LC}=k^+/P^+$ and the Nachtman variable $x_N$ become
\begin{equation}
x_{LC}=\frac{x}{x_N} \left( 1 + \sqrt{1+\frac{4k_{\perp}^2}{Q^2}} \right), \quad 
x_N = 1+\sqrt{1+\frac{4M^2x^2}{Q^2} },
\end{equation}
where  $k_\perp=|\kt|$  is the parton transverse momentum. The scattered quark 
momentum $k'$ is constructed using $k'= k+q$ (see Fig.~\ref{kinem}).
Further, $\phi_k$ is the initial quark  azimuthal angle,
$z_{LC}=P_h^+/k^+$ is the light-cone fraction of the quark momentum carried by the resulting hadron in the $(\tilde x, \tilde y, \tilde z)$-system \cite{Anselmino:2005nn}, where $\tilde z$ is aligned along the scattered quark $k'$.
The final hadron momentum is constructed using,
\be
P_{h\tilde x}=p_{\perp}\cos(\tilde \phi),\quad P_{h\tilde y}=p_{\perp}\sin(\tilde \phi),\quad P_{h\tilde z} = z_{LC} k^\prime_0 - \frac{p^2_{\perp}+M_h^2} {4 z_{LC} 
k^\prime_0}
\ee
where $\tilde \phi$ is the angle between quark and hadron planes, and  $\phi_h$ is the angle between leptonic and hadronic planes according to the Trento convention and $P_{h\perp}$ is the final hadron transverse momentum \cite{Bacchetta:2004jz}.  The final hadron SIDIS variables $\phi_h$, $P_{h\perp}$ and $z$ are  calculated after event generation.
Here we should note, that theoretical or phenomenological distribution and fragmentation functions are  expected to be in the light cone coordinate system 
(see Eq.~\ref{FDALL}).   Motivated 
by the fact that $x_{LC}\simeq x$ and $z_{LC}\simeq z$ 
is a widely used approximation in global fitting, 
 the unpolarized and helicity TMDs are then 
$f_{1,q}(x,\kt^2)$ and $g_{1L,q}(x,\kt^2)$,  
and $D_{1,q}(z,\pt^2)$ is the unpolarized
fragmentation function.  In our Monte Carlo generator  we adopt the parton kinematics
in~\cite{Anselmino:2005nn,Boglione:2011wm} with the additional 
requirements, that the kinematics of the initial and final parton momenta are
kept exact~\cite{Collins:2007ph},   and the
nucleon mass  is not set to zero.

 Finally we note, the Jacobian becomes unity if $k_\perp^2 /Q^2$ corrections are neglected and thus, the usual parton model expression can be recovered in this approximation from Eq.~(\ref{FDALL}).

An 
 interesting question concerns the validity of the 
the parton model and the generalized parton model at the relatively low beam energies available in experiments today.
The 
parton model is an approximation that assumes certain components of the intrinsic parton momenta are suppressed for large beam energies and can thus be integrated out from the distributions. This becomes apparent in Eqns.~(\ref{conv0}) and (\ref{conv}), where the delta function in the $\pm$ components of parton momenta  decouple from transverse momentum resulting in 
a  delta function 
in only the two transverse dimensions.  
An explicit four-momentum conservation law embedded in the formula of the cross section is thus lost.   A  particularly striking consequence that  one observes is that there is no explicit mechanism that prevents events at values of $\Ph$ larger than allowed by the finite beam energy. Naturally, the lower the beam energy becomes, the more serious the inaccuracies of the   parton model have to be taken.  On the other hand, 
the ``fully differential'' cross section Eq.~(\ref{FDALL}) of the generalized 
parton model allows us to include in our Monte Carlo both
transverse momentum and the
 physical energy and momentum phase space constraints.
We used the widely accepted parton model approximation of setting the initial 
parton on-shell (assumption that virtual photon interacts with an on-mass shell quark)\footnote{The confined quark has a non-zero virtuality. Such effects in Monte Carlo generators have 
been studied in Ref.~\cite{Collins:2005uv}.}.
  But it is important to 
emphasize    
 that the approximations we have made,  which are consistent
with a generalized parton model framework, enable  us to
implement a 
Monte Carlo that 
 incorporates the correct phase space momentum constraints and 
satisfies the requirements we outlined in this section.

Thus, our Monte Carlo simulation allows us to take the factorized form of 
the generalized parton model cross section Eq.~(\ref{FDALL}) as a basis and 
then to impose 
four-momentum conservation for the partons according to Fig.~\ref{kinem},
 assuming the initial  quark
 is on-shell with non-zero mass. 
We also take a non-zero target mass into account. 
This procedure does not necessarily lead to a more accurate description of 
the underlying physics, because it still rests on the simplified picture 
of the generalized parton model and involves the approximation of an on-shell 
quark. Nonetheless, implementing these modifications can   give us 
an indication for the magnitude of the uncertainties resulting from the 
aforementioned kinematic approximations in the parton model.

Note that our goal is to study the  applicability of 
Bessel weighting  to experimental data, for which we explicitly need $k_{\perp}$ and $p_{\perp}$ dependences in the 
Monte Carlo generator. 
Alongside with this goal it is interesting to investigate 
how well the approximations of the simple parton model are justified in the current relatively low energy experimental set-up. One would expect that if approximations that lead to the parton model expressions for structure functions are justified, then the generalized parton model expression would not
spoil this approximation numerically. On the other hand if the generalized parton model gives notably different results with respect to a naive parton model, one would expect that kinematics of the experiment does not allow a certain type of approximations and the theoretical/phenomenological description should be improved.

Ultimately the comparison with experimental data will allow us to 
address these questions.  In the mean time in Section~\ref{kincorr} we will study some of 
these issues using our Monte Carlo generator based on the generalized parton model. This will allow us to explore  the validity of certain kinematical approximations and also to understand how parameters of the implemented distributions are different from extracted distributions. Once we have control over these issues in the kinematics of low energy experiments we will
also compare in Section~\ref{Cahn}, our results with data from HERMES experiment as an illustration of possible effects. The applicability of the Bessel weighting technique and resulting uncertainties is a separate issue and will be addressed in Section~\ref{BWALL}.

In the Monte Carlo  generator software, we used the general-purpose, self-adapting event generator,  Foam \cite{TFOAM}, for drawing random points according to an arbitrary, user-defined distribution in  $n$-dimensional space.

\subsection{\label{kincorr}Kinematical Distributions }

 Implementing the Monte Carlo,  we generate kinematical distributions in  $x$, $z$, $k_{\perp}$,  and $p_{\perp}$  of SIDIS events  
for several model inputs of TMDs. These distributions are then used 
to check  the consistency of dependence of extracted quantities under  different model assumptions, including,  for example   Gaussian and non-Gaussian distributions in  transverse momentum.

In case the dependence is assumed to be a Gaussian, $x$ and $z$  dependent widths are assumed, so that TMDs  take the following form, 
\bea
f_{1}(x,\kt^2)&=&f_1(x)\frac{1}{\la k^2_{\perp}(x)\ra_{f_1}}
\exp\left(-\frac{\kt^2}{\la k^2_{\perp}(x)\ra_{f_1}}\right)\, ,\quad
\label{gdf}\\
g_{1L}(x,\kt^2)&=&g_{1L}(x)\frac{1}{\la k^2_{\perp}(x)\ra_{g_1}}\exp\left(-\frac{\kt^2}{\la k^2_{\perp}(x)\ra_{g_1}}\right)\, ,
\label{gdg}\\
D_{1}(z,\pt^2)&=&D_1(z)\frac{1}{ \la p^2_{\perp}(z)\ra}\exp\left(-\frac{\pt^2}{\la p^2_{\perp}(z)\ra}\right)\, , 
\label{gff}
\eea
where $f(x)$ and $D(z)$ are corresponding collinear parton distribution and fragmentation functions
  and the
widths are $x$ and $z$ dependent functions.
In our studies we  adopt  the modified Gaussian  distribution functions 
and fragmentation functions  from Eq.~(\ref{gdf})-(\ref{gff}), in which $x$ and $k_\perp$ dependencies are 
inspired by  AdS/QCD results \cite{SJBrodskyWarsaw2012,deTeramond:2008ht}, with $\la k^2_{\perp}(x)\ra =C\, x(1-x)$
and $\la p^2_{\perp}(z)\ra =D\, z(1-z)$, where the constants $C$ and $D$ may be different for different flavors and polarization states (see for example~\cite{Signori:2013mda}).
 Similarly such non-factorized  $x$,$k_\perp$  distribution functions are  
also suggested by the diquark spectator model \cite{Gamberg:2007wm} and the NJL-jet model \cite{Matevosyan:2011vj,Matevosyan:2013eia}.

For the $x$ and $z$ dependence in Eqs.~(\ref{gdf}) and (\ref{gff}) 
we use the parametrizations, $f_1(x) = (1-x)^3\, x^{-1.313}$, 
$g_{1L}(x) = f_1(x)\, x^{0.7}$, and   $D_1(z) = 0.8\, (1-z)^2$, using input values  $C=0.54\, {\rm GeV}^2$ and 
 $D=0.5\, {\rm GeV}^2$. 
We also assume that $\la k_\perp^2\ra_{g_{1L}}=0.8\, \la k_\perp^2\ra_{f_1}$; this assumption is consistent with lattice studies \cite{Musch:2010ka}  and experimental measurements \cite{Avakian:2010ae}.
 
As an example of a  non-Gaussian $k_\perp$ distribution we implement the following one  inspired by the shape of the resulting distribution in the light-cone quark model \cite{Pasquini:2007iz,Pasquini:2009eb}  
 \be
f_{1}(x,\kt^2)=f_1(x)/\left(  1 +20.82\  k_\perp^2+126.7\ k_\perp^4+1285\ k_\perp^6 \right)\, .
\label{powerkt}
\ee
where the coefficients for $g_{1L}(x,\kt^2)$ are chosen such that
effectively $\la\kp^2\ra_{g_{1L}}/\la\kp^2\ra_{f_1}=0.8$.

\begin{figure}[t]
\begin{minipage}[t]{0.47\linewidth}
\begin{center}
\includegraphics[width=18pc]{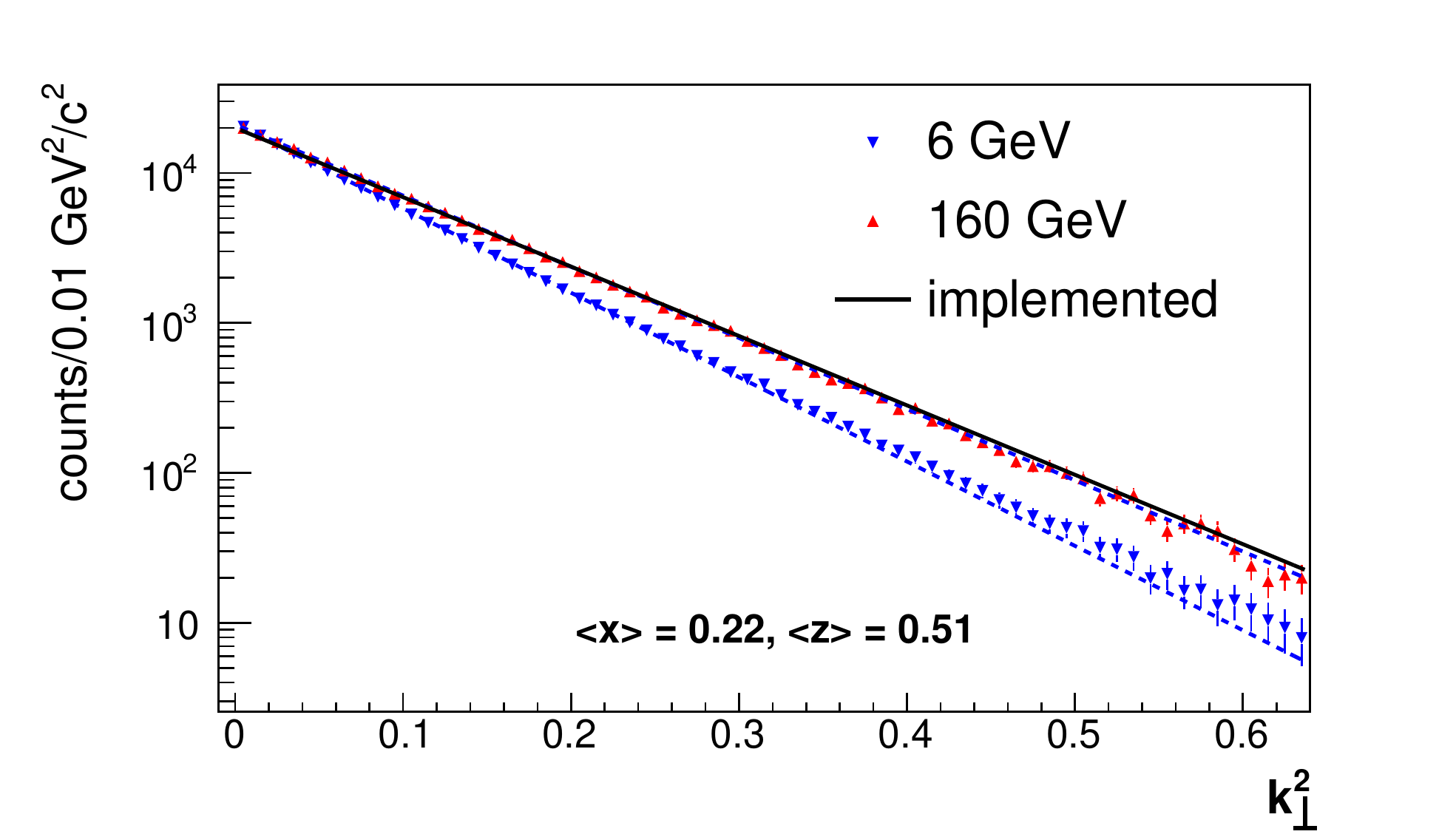}
\end{center}
\begin{center}
\caption{\scriptsize
(Color online) The solid line is the Gaussian input distribution implemented
 using Eq.~(\ref{gdf}), with red triangles coming  from the Monte Carlo  at $160~\rm GeV$ initial lepton energy, blue triangles coming from the Monte Carlo  at $6~\rm GeV$. The dashed line represents the fit to  the Monte Carlo  distributions which returned values  of $C= 0.527\, {\rm GeV^2}$ and $C=0.444\, {\rm GeV^2}$ at $160~\rm GeV$ and $6~\rm GeV$ respectively.}\label{ktfit} 
\end{center}
\end{minipage}
\hspace{2pc}%
\begin{minipage}[t]{0.47\linewidth}
\begin{center}
\includegraphics[width=18pc]{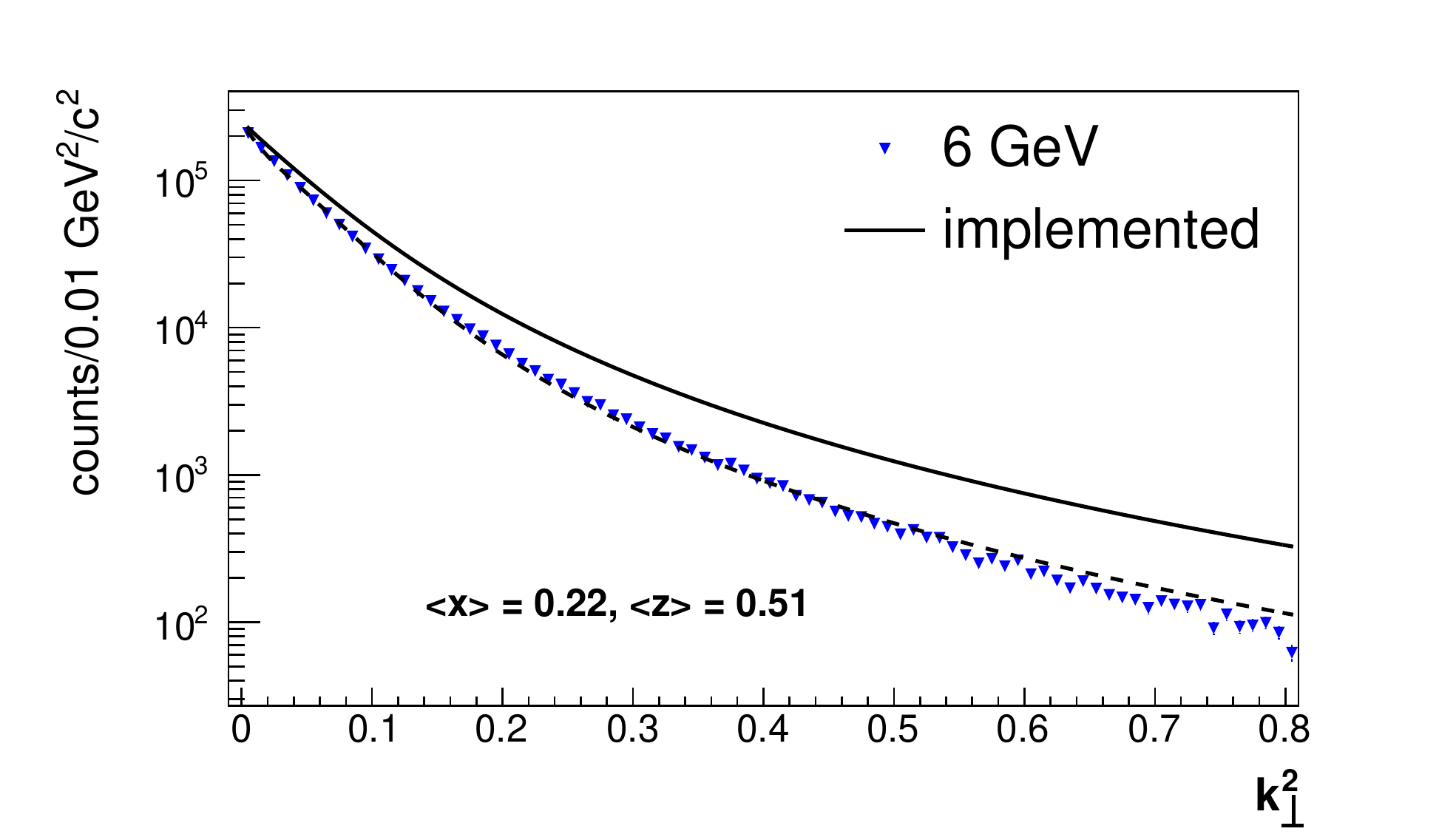}
\end{center}
\begin{center}
\caption{\scriptsize (Color online) The solid line is the  implemented
  non-Gaussian distribution using Eq.~(\ref{powerkt}), 
with $\la\kp^2\ra=0.084\, {\rm GeV}^2$, and the 
dashed curve represents the fit to the Monte Carlo 
distribution 
with the 
value of $\la\kp^2\ra=0.064\, {\rm GeV}^2$ at $6\, {\rm GeV}$ initial 
lepton beam energy. The available phase space dictated by four momentum
conservation results in a deformation of  the input distribution. }
\label{3rdorderPOLfitwithpl3}
\end{center}
\end{minipage}
\end{figure}

We then generate events  using the cross section from Eq.~(\ref{FDALL}) for both Gaussian and non-Gaussian initial distributions respectively,   and we
display the
resulting transverse momentum distributions in
 Figs.~\ref{ktfit} and \ref{3rdorderPOLfitwithpl3}.  
Note (as stated earlier) that the generator we construct is  implemented with 
on-shell initial partons  with four momentum conservation imposed. While this 
choice is not compulsory we adopt it as it allows us to fully reconstruct 
kinematics for a given event. At the same time, the limitations due to available 
phase space integration will modify the reconstructed distributions with respect to the input distributions.  
We analyze the effect of the available phase space in the Monte Carlo   
on the average  $\la\kp^2\ra$  
for finite beam energies as a function of $x$ by calculating the 
effective $\la\kp^2\ra$ from the following formula,
\bea
\la\kp^2(x)\ra = \frac{\int d^2\kt \kp^2 d\sigma_{MC}}{\int d^2\kt d\sigma_{MC}}=
\frac{\sum_{j=1}^N  \kpj^2}{N}\,  ,
\label{kperpav}
\eea 
where the index $j$ runs over the $N$ Monte Carlo  generated events.
Note, $d\sigma_{MC}$ is the  cross section of the Monte Carlo 
simulation, that is Eq.~(\ref{FDALL}),  modified by imposing 
the four momenta conservation and on-shell condition for initial quark.

 Indeed in Figs.~\ref{ktfit} and \ref{3rdorderPOLfitwithpl3} we find 
when comparing the Monte Carlo generated events with   the  {\it input}  distributions, using
 Eq.~(\ref{gdf}) and  Eq.~(\ref{powerkt}),  shown as solid black curves
 for a given $x$, that the larger $\kp$ values of the Monte Carlo 
events (red triangles up, 160 GeV beam energy, and blue triangles down, 6 GeV beam energy)  are suppressed due to the
available phase space imposed by both the finite beam energy, and  four momentum conservation in  the Monte Carlo.
  The fit of the Monte Carlo distributions for the modified Gaussian model
are shown as  dashed lines
 displayed in Fig.~\ref{ktfit}. 
  They return the fitted values $C=0.527\, {\rm GeV}^2$ and $C=0.444\, {\rm GeV}^2$ for the
$160\, {\rm GeV}$ and $6\, {\rm GeV}$ Monte Carlo simulations respectively.
In Fig.~\ref{3rdorderPOLfitwithpl3} we study the effect of the
non-Gaussian distribution Eq.~(\ref{powerkt}). Integrating 
 Eq.~(\ref{kperpav}) over $\kp$ gives a value of 
 $\la\kp^2\ra=0.084\, {\rm GeV}^2$, and the 
dashed curve represents the fit to the Monte Carlo 
distribution with  a value of 
$\la\kp^2\ra=0.064\, {\rm GeV}^2$ for the  $6\, {\rm GeV}$ initial 
lepton beam energy. 

\begin{figure}[h]
\begin{minipage}[t]{0.47\linewidth}
\includegraphics[width=18pc]{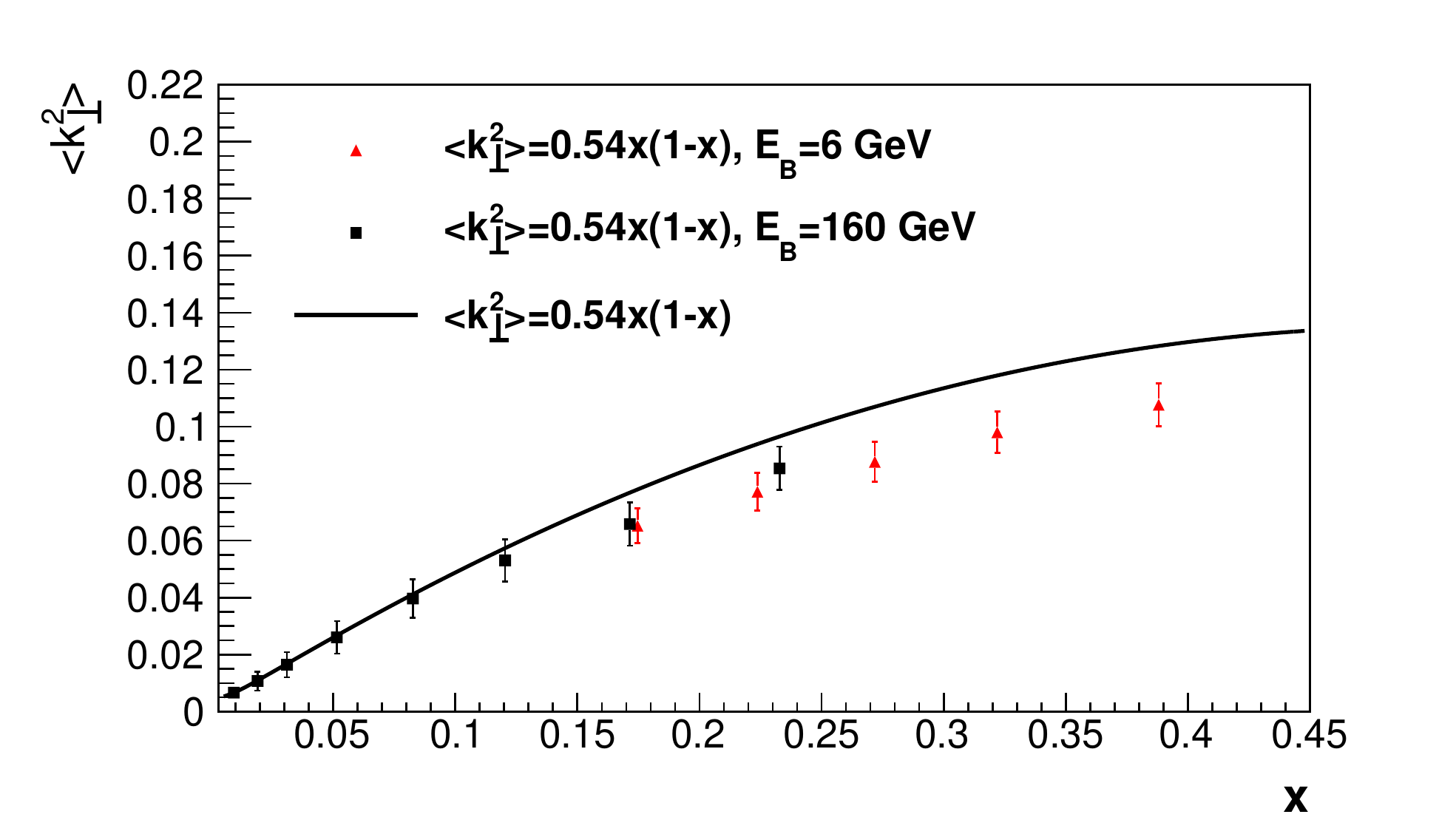}
\caption{\scriptsize
(Color online) $\la\kp^2(x)\ra$ versus $x$ for the unpolarized ($d\sigma^+ + d\sigma^-$ ) cross-section  for $0.50<z<0.52$, for two Monte Carlo runs
with beam energies 6 GeV and 160 GeV, 
 with the modified Gaussian  distribution function 
and fragmentation functions. 
The solid line represents the input function, while 
the Monte Carlo generated  values are black squares for 160 GeV and red triangles for 6 GeV.}\label{kTvsxB} 
\end{minipage}
\hspace{2pc}%
\begin{minipage}[t]{0.47\linewidth}
\includegraphics[width=18pc]{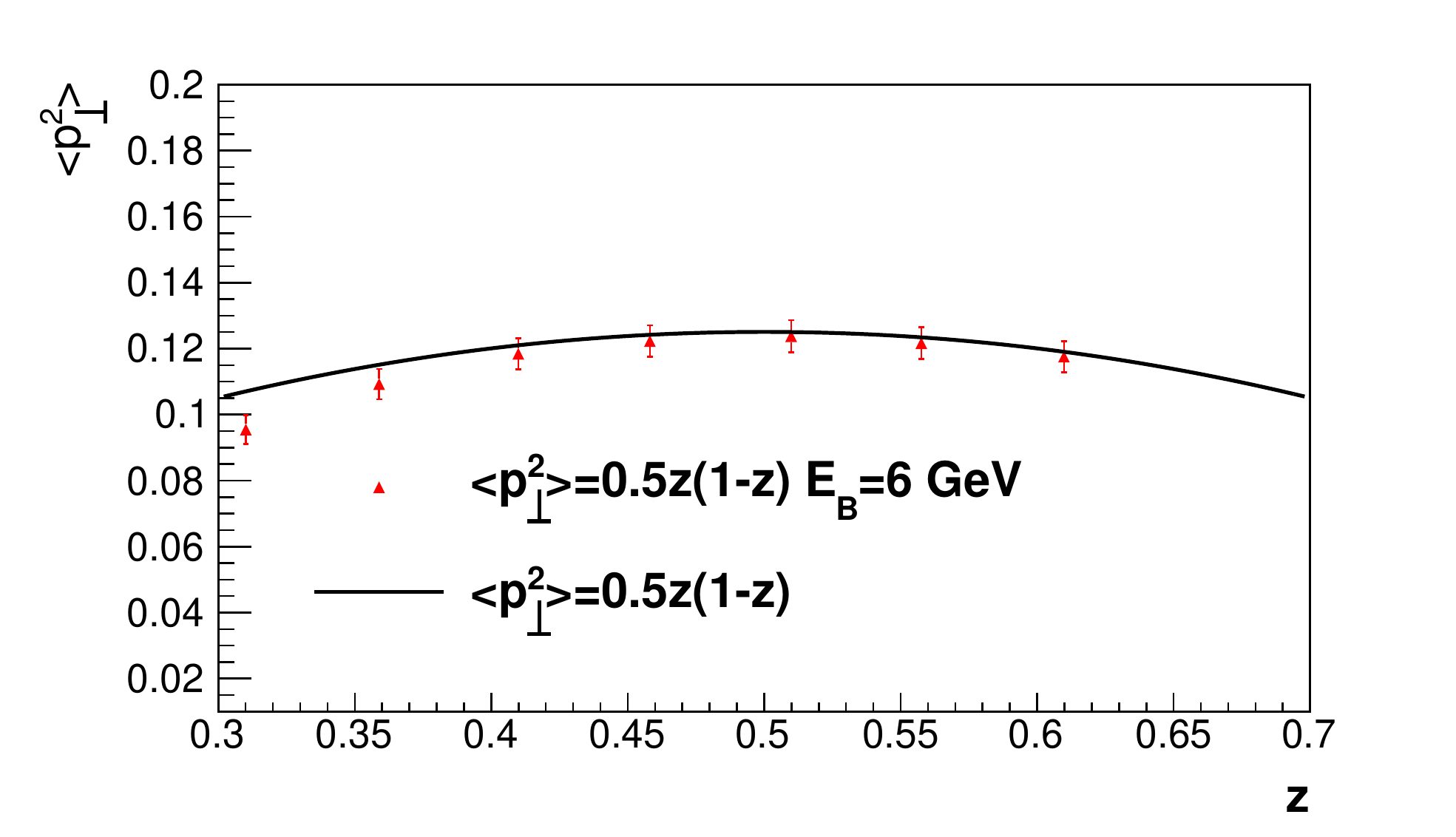}
\caption{\scriptsize (Color online) $\la p^2_{\perp}(z)\ra$ versus $z$ for the unpolarized ($d\sigma^+ + d\sigma^-$ ) cross-section for the $6~{\rm GeV}$ beam for $0.20<x<0.25$ from the Monte Carlo  with the modified Gaussian distribution  and fragmentation functions as compared to the analytic result using Eq.~(\ref{FDALL}) 
and the input distributions.  The solid line represents the input
function, while 
the Monte Carlo generated  values
 are the  red triangles for 6 GeV.}
\label{pTvsZ}
\end{minipage} 
\end{figure}

In Fig.~\ref{kTvsxB}.    
the average $\la\kp^2\ra$ versus $x$ from the Monte Carlo  
for different incoming beam energies, for $0.5 < z < 0.52$,  is presented.  
For the modified Gaussian  distribution function  with the 
input value $\la\kp^2(x)\ra = 0.54\, x\, (1-x)\, {\rm GeV}^{2}$, 
the suppression of the generated  $\la\kp^2(x)\ra$ compared to
input distributions (solid line) is greater for the lower beam energy.  
In Fig.~\ref{pTvsZ} the  constraints  of 
four momentum conservation  also affect  
the  $\pp^2$ distributions,  which in turn also affect the observed $P_{h\perp}$ distribution.

The systematic deformation of the extraction of the TMDs in momentum space due to the 
kinematic constraints has been studied in detail using our fully differential
Monte Carlo.  We conclude this section with the general observation
  that  imposing four momentum conservation 
in the event generator  effectively modifies the initial distributions 
due to the limitations of the  available phase space in the  generator.   
 This deformation is  more pronounced
 at lower energies or $Q^2$. A shift of a few percent is visible for  
$160\, {\rm GeV}$  incoming lepton beam energy,  while for the lower  $6$ GeV 
 energy  the {\it effective}  $\la\kp^2\ra$ is lower than the input 
value by approximately $\sim$ 20\%. 

 At this point let us comment on the applicability of the description of experimental data in terms of a simple parton model. As we can see
from Figs.~\ref{ktfit}, and \ref{kTvsxB}, the
results are  consistent for large energy (160 GeV) while they exhibit a 
significant shift at lower energy (6 GeV)  for the same 
 input parameters.
In fact the kinematical corrections due to imposing  four-momentum conservation
 and  target mass corrections, grow at lower energies as one would  expect.  At the same time, a 20\% correction is well  within the expected accuracy of the parton model approximation; remember that one usually neglects $\kp^2/Q^2$ corrections, and one would expect that after inclusion of  such  corrections one can 
achieve a better  quantitative description of the data. 
In the next  section we present the outcome of the Monte Carlo
 compared to experimental data from  HERMES, 
and discuss its relevance.

\subsection{\label{Cahn}The  Cahn effect in the Monte Carlo Generator}

As an example of an application of our constructed 
 Monte Carlo we present a study of the 
 Cahn effect~\cite{Cahn:1978se,Cahn:1989yf} contribution to the
average  $\la\cos\phi\ra$ moment in SIDIS.   We generate Monte Carlo  events using the following expression for the cross section~\cite{Anselmino:2005nn}, 
\bea
\frac{d \sigma}{ dx dy dzd^{\, 2}{\pt}d^{\, 2}{\kt} } =K(x,y)J(x,Q^2,\kt^2) \sum_a f_{1,a}(x,\kt^2)D_{1,a}(z,\pt^2)  \frac{\hat s^2 + \hat u^2}{Q^4}
\label{FDC}
\eea
where $\hat s=(l+k)^2$ and $\hat u = (k-l')^2$ (see Fig.~\ref{kinem}).
As stated above, in the Monte Carlo  we impose four momentum conservation with target mass corrections. 
 
\begin{figure}[ht]
\centerline{\includegraphics[height=2.0in]{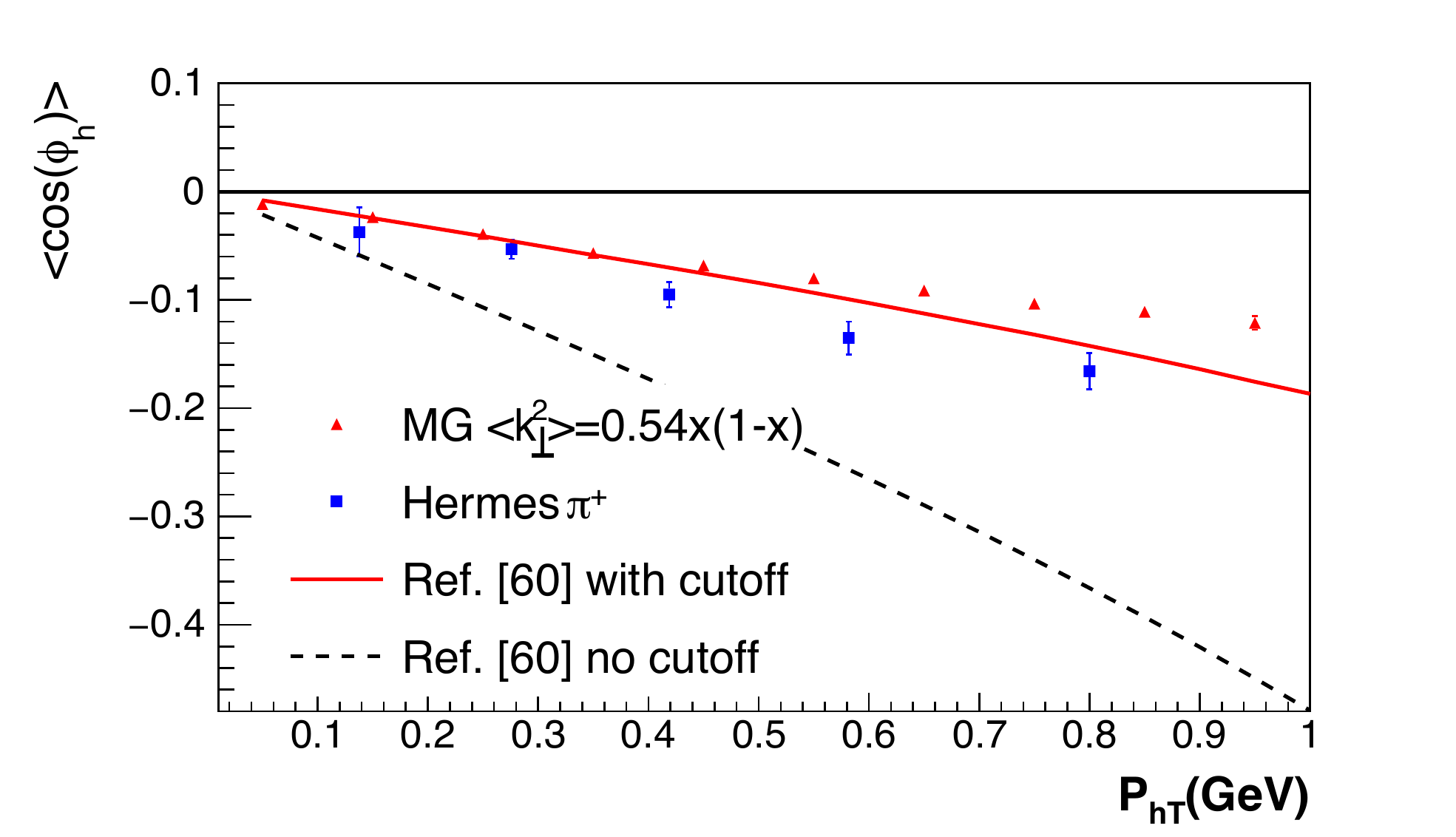}}
\caption{\scriptsize (Color online) The Cahn contribution in $\la\cos(\phi_h)\ra$ for $\pi^+$ from the modified Gaussian (red triangles denoted as MG 
in the figure) PDFs  using  Eq. \ref{gdf}  is presented for HERMES kinematics in comparison with  Ref. \cite{Boglione:2011wm} (red solid and 
black dashed lines) and published HERMES data~~\cite{Airapetian:2012yg} (blue squares).}  
\label{CahnSBA_HERM}
\end{figure}

In Fig.~\ref{CahnSBA_HERM} we present output from the Monte Carlo  using the 
non-factorized Gaussian distribution function  and 
fragmentation function (Eqs.~(\ref{gdf}) and (\ref{gff})).
We also compare our results to the HERMES data~\cite{Airapetian:2012yg}, and
  Ref.~\cite{Boglione:2011wm}.   The dashed line in  Fig.~\ref{CahnSBA_HERM}
  represents the naive parton model result without any kinematical constraint on parton momenta while  the  solid line results from
  performing the computation   with the kinematical constraints.
 One can see that taking into account these constraints is  
important for a description of the experimental data within this model.
It is clear that the results of our Monte Carlo are comparable to that of~\cite{Boglione:2011wm} and close to HERMES data~\cite{Airapetian:2012yg}.  
For the red triangles,  we used $\la\kp^2\ra=0.54\; x(1-x)$ 
and $\la\pp^2\ra=0.5\; z(1-z)$ GeV$^2$. As one can see for HERMES kinematics the modified Gaussian TMDs reduces
 the contribution of the Cahn effect contribution to the   $\la\cos(\phi_h)\ra$
moment. 
In Ref.~\cite{Boglione:2011wm} this effect is achieved by imposing a 
so-called direction cut (that the quark moves in the forward direction with respect of the proton). In this Monte Carlo there are two main factors that 
modify the distribution; the four-momentum conservation 
and $x(z)$ dependent values of $\la\kp^2\ra\, (\la\pp^2\ra)$. 
One might expect that the Jacobian in Eq.~(\refeq{FDC}) plays a major role modifying transverse shape of resulting cross section, however we checked that it is not the case. 
The most important effect comes from taking into account kinematical constraints on parton momenta. One would conclude that taking these corrections into account is  important for reliable analysis of experimental data.

 In the next Section we apply the Bessel weighting formalism for 
the double longitudinal 
spin asymmetry in semi-inclusive deep inelastic 
scattering to data from our  Monte Carlo generator.

\section {\label{BWALL} Bessel Weighted 
Double Spin Asymmetry }
\label{sec:extract}

In this Section, we present an extraction of the Bessel weighted 
double longitudinal spin asymmetry 
in $\btu$ space. 
We also carry out a study of the accuracy of such an extraction.  
We use the dedicated fully differential SIDIS  single hadron Monte
Carlo to generate events based on the input TMDs.  For simplicity we perform this comparison in a one flavor approximation.

\subsection{ Results from the Monte Carlo }
\label{sec:BWALLres}

\begin{figure}[b]
\centerline{\includegraphics[height=3.5in]{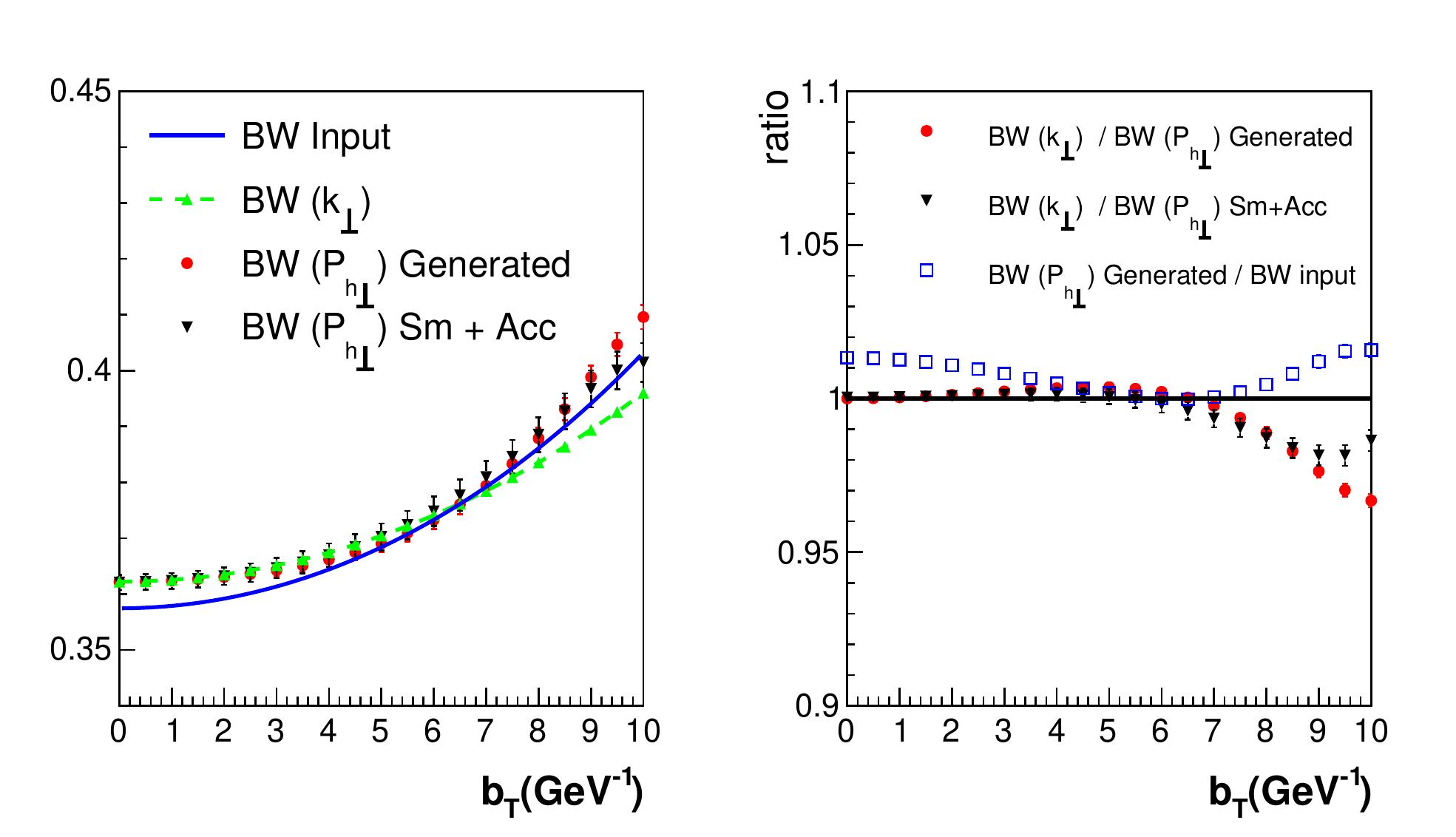}}
\caption{ \scriptsize (Color online) Left panel:
The ratio of Fourier transforms $\tilde g_{1L}/\tilde f_1$ 
and the   Bessel weighted 
asymmetry $A^{J_{0}(b_T\qtu)}_{LL}$ plotted versus $b_T$. The solid curve (blue) is the Fourier transform of the input to the Monte Carlo given by  Eq.~(\ref{tildall}), the red points are    
 generated  Monte Carlo  events using Eq.~\ref{spma1},  and  triangles down (black) represent results of Monte Carlo  events after experimental smearing and acceptance at 
$\la x\ra=0.22$, and $\la z\ra=0.51$. The  triangles up with dashed curve (green) are results of the Monte Carlo without inclusion of fragmentation 
functions (see text for discussion of errors).
Right panel: Ratios that represent the accuracy of our results.
}
\label{BWPhTcorr}
\end{figure}

\begin{figure}[ht]
\centerline{\includegraphics[height=3.5in]{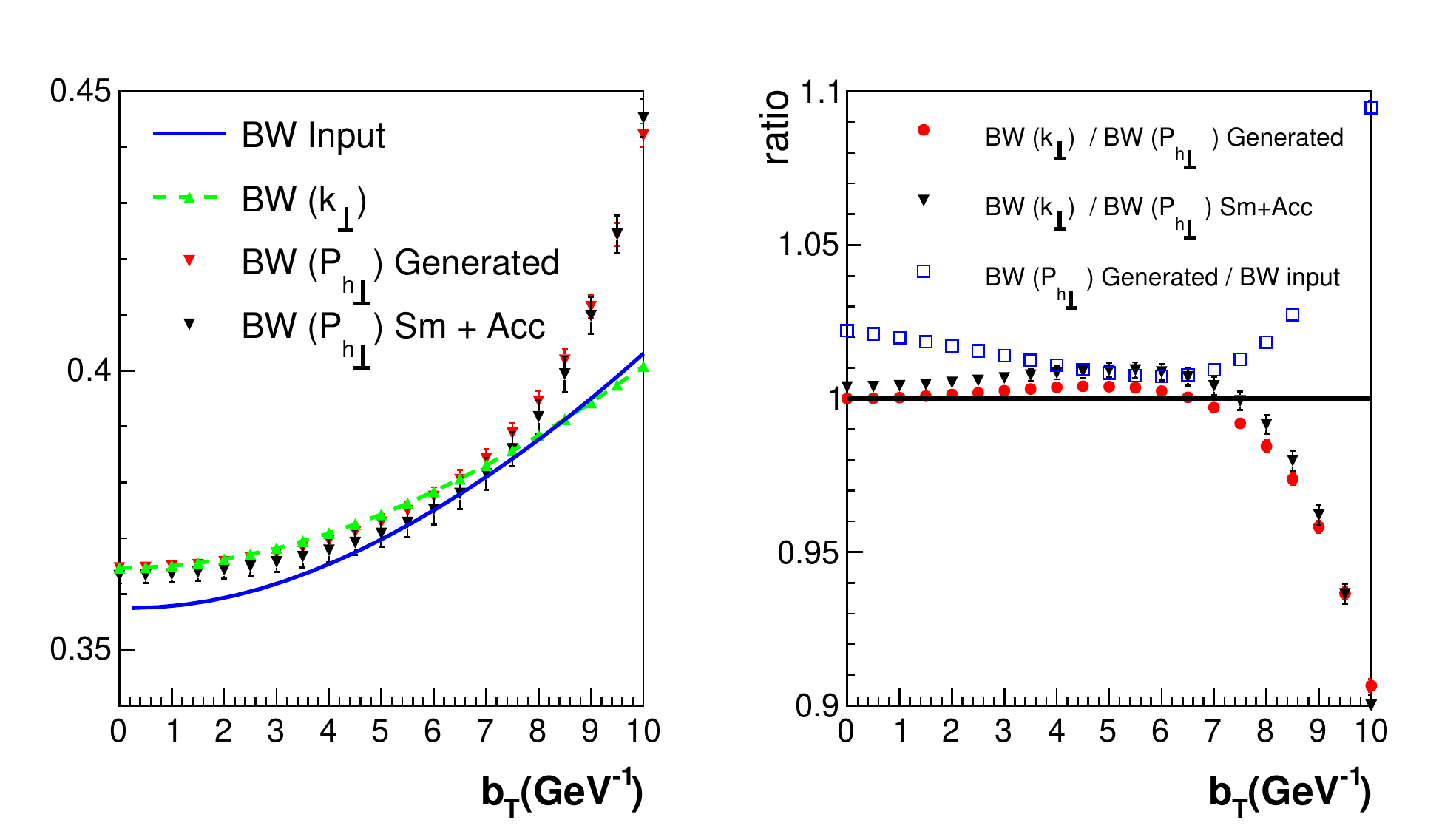}}
\caption{\scriptsize (Color online) The same as in Fig.~\ref{BWPhTcorr} but here from the power-law tail distribution function based on the Monte Carlo
(see text for discussion of errors).}
\label{BWPhTcorrg1f1}
\end{figure}

The Monte Carlo  generated events  are  used like experimental events
to extract both the
Bessel weighted asymmetry $A^{J_{0}(b_T\qtu)}_{LL}$, and the ratio of the Fourier transform of $g_{1L}$ to $f_1$, using the Bessel weighting method described in~\cite{Boer:2011xd}. The results  
are then compared to the Monte Carlo  input.  The Bessel moments are extracted from  the Monte Carlo  with $6 ~{\rm GeV}$ beam energy using both the  modified Gaussian type of functions (see Eqs.~(\ref{gdf})-(\ref{gff})) and power law-tail like function (see Eq.~(\ref{powerkt})).

The numerical results of our studies are summarized 
 and displayed  in Figs.~\ref{BWPhTcorr} and \ref{BWPhTcorrg1f1}  for the 
modified Gaussian distribution function and for the power law-tail like  distribution function inputs respectively.   
In the left panel of Fig.~\ref{BWPhTcorr}
we show the Bessel-weighted asymmetry 
versus   $b_T$.   The blue curve labeled  ``BW Input'',  
is the asymmetry calculated analytically using the right hand side of Eq.~(\ref{tildall})
and the Fourier transformed input distribution functions (one can compare this with 
the model calculation in Ref.~\cite{Lu:2012ez}).  

 We now compare various distributions
generated from the Monte Carlo. We plot 
 the generated distribution using  Eq.~(\ref{spma1}) (full red points) labeled
``BW($\Ph$) Generated'', 
and  the black triangles  labeled 
``BW($\Ph$) Sm + Acc'', which  represents the same extraction after  experimental 
smearing and acceptance. 
For this we  use the  CLAS spectrometer~\cite{Mecking:2003zu},
which is a quasi-$4\pi$ detector, comprised of six azimuthally symmetric detector arrays, and uses a toroidal field to bend charged particles. Particle momenta and scattering angles were measured with a drift chamber tracking system with a relative accuracy of $0.3\%$ to $2\%$ in momentum, and about 3 mr in angle and
 with less than $1\%$  
momentum resolution in the presented bin
$\la x\ra=0.22$, and $\la z\ra=0.51$.

Next we consider the Fourier transform 
ratio $\tilde g_{1L}$ to $\tilde f_1$,
the (green) curve with triangles up labeled ``BW($\kp$)'' obtained from
 numerically Fourier transforming the $\kp$ distributions from the Monte 
Carlo generator on an event by event basis (see Eq.~(\ref{eq-bttransf})),
\bea
\sqrt{1-\varepsilon^2}\frac{\tilde g_{1L}(b_T)}{\tilde f_1(b_T)}
 &=& 
\frac{\displaystyle\sum_{{j}}^{N^{+}}\,  J_0(\btu {\ktpj}) 
-
\sum_{{j}}^{N^{-}}\,  J_0(\btu {\ktmj})}
{\displaystyle\sum_{{j}}^{N^{+}}\,  J_0(\btu {\ktpj})
+\sum_{{j}}^{N^{-}}\,  J_0(\btu {\ktmj})}\, .
\label{spma2}
\eea

This quantity corresponds to the right hand side of  Eq.~(\ref{tildall}) in the one flavor approximation, where
the  fragmentation functions are expected to cancel out. For the Monte Carlo generated events, this cancellation is only an approximation, leading to the deviation between the red (or black) points and the green curve at large $\btu$. 
The reasons for the imperfect cancellation are discussed in section \ref{sec:intrp}.

In order to quantitatively assess  the  deviation between the
curves in the left panel of Fig.~\ref{BWPhTcorr}, 
we plot  ratios of these values (see right panel). 
The red points represent the deviation from unity 
that is due to the imperfect 
cancellation
of the fragmentation function.  The black
triangles represent the same after experimental smearing and acceptance
are taken into account. 
Finally the open blue squares represent the 
deviation between 
the  analytic result from the input distributions 
and the Monte Carlo generated events,  Eq.~(\ref{spma1}).

The error bars in $b_T$ space for each point give the statistical standard deviation. For each $b_T$ point,  
the statistical error bars are calculated from ${}\sim 250 {\rm k}$
independently generated events in momentum space
(see Appendix~\ref{errors} for more details on error calculations). However,
 if we use the same data set to integrate over  $P_{h\perp}$ for all $b_T$ points, the errors in $b_T$ space are correlated, which needs to be taken into account in the error analysis of any global fit to the data points.
To circumvent this problem, we used different Monte Carlo  samples for each $\btu$-point. 
 In our numerical simulation we can afford to do that, because we can generate events  copioulsy.

The same idea is applicable for future experiments at Jefferson Lab 12~\cite{Dudek:2012vr}, RHIC Spin ~\cite{Aschenauer:2015eha} and Electron Ion Collider~\cite{Accardi:2012qut} that will deliver events in great abundance.  
In that case, one can divide the data into subsets, and take independent subsets of data for each value of $b_T$ to calculate the asymmetry.
As a rule of thumb, if an experimental analysis can 
afford to have 5-10 bins in $\Ph$ , then we expect there to be enough data to split it up for independent analyses at 5-10 values of $b_T$.
However, if events are scarce, we need to find a
way to carry out the error analysis using the same dataset for all values 
of $b_T$. An obvious way to do that might be to perform a generalized least squares fit based on the correlation/covariance matrix. Be aware, however, that using the inverse sample covariance matrix in a fit can be an unstable approach, for two reasons. First of all, the sample covariance matrix becomes a poor estimate of the true covariance matrix when the number of fit points is not many times smaller than the number of experimental observations, i.e., $N^\pm$. In particular, the sample covariance matrix may be (close to) singular. Secondly, any systematic effects in the data that have been neglected in the model of the fit function may be amplified strongly by the inverse covariance matrix. Therefore, it often turns out to be more practical and stable to perform a weighted least squares fit, ignoring all correlations. This is a valid approach as long as we do not ignore the correlations in the subsequent error analysis of the resulting fit parameters. 
Resampling techniques provide simple yet powerful solutions to perform the correlated error analysis and are  very popular in lattice QCD. We can take advantage of them in our situation as well. As explained in greater detail in Appendix \ref{errors} and \ref{bootstrap}, the sums $\tilde S^\pm =  \sum_{{j}}^{N^{\pm}}\,  J_0(\btu \qtpmj)$ in Eqn. (\ref{spma1}, \ref{spma}) follow so-called \textit{compound Poisson distributions} \cite{adelson1966compound}. Therefore, we can make use of the variant of the bootstrap method proposed in Ref. \cite{Bohm:2013gla}, which we describe in the context of Bessel weighted asymmetries in Appendix \ref{bootstrap}.

\subsection{Interpretation of the Results}
\label{sec:intrp}

One primary question addressed in this study is how robust the Bessel-weighting technique behaves under simulating real experimental conditions.  Comparing the round (red) data points with the triangular (black) ones in Figs. \ref{BWPhTcorr} and \ref{BWPhTcorrg1f1}, we see that switching on experimental smearing and acceptance in our simulation does not change the results significantly.     
 Analyzing our MC results with four momenta conservation and target mass correction, we are able to distinguish two effects in the left panels of Figs.~\ref{BWPhTcorr} and \ref{BWPhTcorrg1f1}:

\begin{enumerate}
\item {\bf Solid (blue) curve versus triangular (green) data points: The distributions realized in the MC simulation differ from the input distributions.} 
In the MC, the four-momentum conservation does not allow the variables $\kt$ and $\pt$ in Eq.~(\ref{FDALL}) to be sampled independently over the whole integration range, as it would have to be done to reproduce the unmodified parton model Eqns.~(\ref{conv0}) and (\ref{conv}). 
The actual $\kt$ and $\pt$ distributions realized by the MC differ from the analytic input distributions Eqns. (\ref{gdf})-(\ref{powerkt}) noticeably, especially in their widths. This has already been observed in Fig. \ref{ktfit}. The solid (blue) curve in the left panel of Figs.~\ref{BWPhTcorr} and \ref{BWPhTcorrg1f1} is calculated from the input distributions according to the parton model; the FFs on the right hand side of Eq. (\ref{tildall}) cancel exactly in the single flavor scenario. Thus the solid curve can be compared to the triangle shaped (green) data points, which have also been calculated from a ratio of TMD PDFs, Eq. (\ref{spma2}), albeit with the actual distributions realized in the MC.    

\item {\bf Triangular (green) data points vs. circular (red) data points: inadequacy of the generalized parton model to describe the data.}  In a single flavor scenario, the distribution functions $\tilde D_1^a$ cancel exactly on the right hand side of Eq. (\ref{tildall}). Therefore, there should not be any difference between the  full asymmetry $A^{J_{0}(b_T\Ph)}_{LL}(b_T)$ of Eqs. (\ref{tildall}), (\ref{spma1}) and the ratio of TMD PDFs Eq. (\ref{spma2}). However, we do observe a difference between the circular (red) data points and the triangular (green) data points in the left panels of Figs. \ref{BWPhTcorr} and \ref{BWPhTcorrg1f1}. Again, the four-momentum conservation we have implemented is the reason for the observed difference. Since $\kt$ and $\pt$ are no longer sampled in accordance with Eqns.~(\ref{conv0}) and (\ref{conv}), the right hand side of Eq. (\ref{tildall}) then needs modification~\footnote{Such issues have been discussed
in the context of a Monte Carlo generator 
in~\cite{Collins:2005uv,Collins:2004vq}.}. 
Therefore, we see only an incomplete cancellation of FFs 
for the Monte Carlo events. 
\end{enumerate}

To an experimentalist who is concerned about systematic errors attributed to the observables he or she extracts, the first of the two effects above is not an issue. The purpose of the generalized parton model is to provide a parametrization of the data one observes. Any effect of the underlying scattering mechanism that can be absorbed into the distributions does not contradict the validity of the model. The only concern one might have is that the distributions become beam energy/$Q^2$ dependent, an issue that should be addressed using TMD evolution equations.

On the other hand, the second effect presented above {\it can} be taken as an indication for systematic uncertainties. If, indeed, the physical reality does not generate events in accordance with the functional shape of the generalized parton model, then using the model for the extraction of distributions necessarily involves systematic errors. Again, we point out that it is unclear whether the modifications we have implemented in our MC bring us closer to the physical reality. Nonetheless, the modifications are reasonable and  so we believe they can give us a hint about the order of magnitude of systematic errors from the corresponding approximations in the model. 
One can then estimate that for calculations such as those performed in 
Ref.~\cite{Lu:2012ez},
 systematic errors in the comparison with experimental data for 
$\btu < 6\, {\rm GeV}^{-1}$ are of the order of a few percent.  For the data with $\btu > 6\, {\rm  GeV}^{-1}$, the effects of four-momentum conservation (difference between red and green points) becomes more pronounced, and a fit of data using the parton model, i.e., without manifest four-momentum conservation, therefore becomes less accurate.

\section{Conclusions}
\label{sec:conc}
We have presented the first studies of Bessel-weighted asymmetries using a multi-dimensional Monte Carlo generator based on the fully differential cross section for TMD studies using the tree level parton model 
\cite{Anselmino:2005nn}. 
Two models have been used in the simulation; a 
modified Gaussian and a power law tail,  for the distribution and fragmentation functions. The Bessel-weighted sums of
double polarization observables, in particular, provide access to transverse momentum dependencies of partonic distributions $f_1$ and $g_{1L}$.
Bessel-weighted asymmetries (described in \cite{Boer:2011xd}) have been extracted from the generated Monte Carlo  events and studies of systematic uncertainties  have been performed. 
 We observe a few percent  systematic offset of the Bessel-weighted  asymmetry obtained from Monte Carlo  extraction compared to input model calculations, 
which is due to the limitations imposed by the energy and momentum
conservation at the given energy/$Q^2$.

We find that the Bessel weighting technique provides a powerful and 
reliable tool to study the Fourier transform of TMDs with controlled 
systematics due to experimental acceptances and resolutions with different  
TMD models inputs.  
We plan to expand our studies with more advanced parton shower and fragmentation mechanisms, as well as to include nuclear modifications in our Monte Carlo and extraction procedure.

A Monte Carlo generator including spin-orbit correlations, quark-gluon interactions and correlations between the current and target fragmentation region, which
 is  applicable in a wide range of kinematics, will be crucial for both experimental techniques and phenomenology of Fourier transformed TMDs.  Moreover, evolution equations for the distributions are typically formulated directly in coordinate (Fourier) space~\cite{Collins:1981uk,Collins:1981uw,Collins:1984kg,Ji:2004wu,Collins:2004nx, Bacchetta:2008xw,Aybat:2011zv,Collins:2011zzd}. Phenomenological studies can then be performed in this space, see for example~\cite{Konychev:2005iy,Aidala:2014hva}. Thus, the study of the scale dependence of Bessel weighted asymmetries should prove important in studies of evolution of TMDs. 
  For the above stated reasons we propose Bessel weighted asymmetries as clean observables to study the scale 
dependence of \TMDPs and FFs at existing (HERMES, COMPASS, JLab) and future facilities (Electron Ion Collider, JLab 12 GeV).

\section*{Acknowledgments}

This work is supported by the U.S. Department of Energy under Contract 
No. DE-AC05-06OR23177 (H.A., A.P., P.R.), No. DE-FG02-07ER41460 (L.G.), 
 Science without borders young talent program from CAPES (contract number 150324 da CAPES), EU FP7 (HadronPhysics3, Grant Agreement number 283286) (M.A.), and the Italian Istituto Nazionale di Fisica Nucleare (M.A., H.A., E.De-S., M.M., P.R.).  We thank  M. Anselmino, D. Boer, S. Brodsky, U. D'Alesio, D. Hasch, A. Kotzinian, H. Matevosyan,  and S. Melis for useful and stimulating discussions. We would like to thank the referee of this paper for his/her thoughtful comments that helped us to sharpen physics discussion and improve presentation of results.

\appendix
\section{Bessel Weighting\label{Projectbin}}

In this Appendix we  review the Bessel weighting framework, 
and the  procedure to calculate the 
Bessel-weighted asymmetry  
for the longitudinally polarized beam and target, for a given set of  experimental events which
is expressed in  Eq.~(\ref{spma1}).

From Eq.~(\ref{eq:CS_bspace}) the SIDIS cross section written in terms of the Fourier transformed TMD PDFs and FFs~\cite{Boer:2011xd} for the leading twist 
unpolarized and doubly longitudinal polarized structure functions is given by
\bea
\frac{d\sigma}{dx\, dy\, d\psi\, dz\, d\phi_h\, d\Ph^2}&=&K(x,y)\int\frac{d\bp\, \bp}{2\pi}
J_0(\bp\, \Ph )\left(\mathcal{F}_{UU,T}(\bp)+S_{||}\lambda_e\sqrt{1-\varepsilon^2}
\mathcal{F}_{LL}(\bp)\right)\, \nn
\eea 
where $K(x,y)$ is given in Eq.~(\ref{Kfact}) and where 
$\Php\equiv \Ph$ and $|\bt|\equiv\bp$ .

Using the Bessel weighting procedure, 
which in this case amounts to weighting  with $J_0$,
we write the cross section  $\tilde{\sigma}(\Bt)$ in $\Bt$ space,   in terms of the structure functions\footnote{We have
suppressed the dependence on the phase space variables $x,y,z$.}
  $\mathcal{F}_{UU,T}$ and $\mathcal{F}_{LL}$  
\bea
\tilde{\sigma}(\Bt)&=&2\pi\int d\Ph\Ph J_0(\Bt\Ph)\frac{d\sigma}
{dx\, dy\, d\psi\, dz\, d\phi_h\, d\Ph\, \Ph}\nn
&=&2\pi\int d\Ph\Ph J_0(\Bt\Ph)
\int\frac{d\bp\, \bp}{2\pi}
J_0(\bp\, \Ph)
\left(\mathcal{F}_{UU,T}+S_{||}\lambda_e\sqrt{1-\varepsilon^2}
\mathcal{F}_{LL}\right)\nn
&=&
K(x,y)
\left(\mathcal{F}_{UU,T}+S_{||}\lambda_e\sqrt{1-\varepsilon^2}
\mathcal{F}_{LL}\right)\, ,
\label{eq:FTdsigma}
\eea
where the structure functions in $\bp$ space are given by the products
of Fourier transformed TMDs~\cite{Boer:2011xd},
\bea
  \FTStrufu _{UU,T} = x\, \sum_a e_a^2 \tilde f_1^a(x, z^2\btu^2)\tilde D_1^a(z, \btu^2)\, ,  \;
\FTStrufu_{LL}  =  x\, \sum_a e_a^2 \tilde g_{1L}^a(x, z^2\btu^2) \tilde D_1^a(z, \btu^2)\, . 
\eea
Labeling the cross section with $\pm$ for the double longitudinal spin combinations $\Sp\lambda_e=\pm 1$ we have
\bea
\tilde{\sigma}^\pm(\bp)
&=& K(x,y)
\left(\mathcal{F}_{UU,T}\pm\sqrt{1-\varepsilon^2}
\mathcal{F}_{LL}\right)\, .
\label{cspma}
\eea
The Bessel weighted double spin asymmetry is $b_T$ space is,
\bea
 A^{J_{0}(b_TP_{h\perp})}_{LL}(b_T) = \frac{ \tilde \sigma^+(b_T) - \tilde \sigma^-(b_T)}{\tilde \sigma^+(b_T) + \tilde \sigma^-(b_T) }\equiv\frac{\tilde \sigma_{LL}(b_T)}{\tilde \sigma_{UU}(b_T)}=\sqrt{1-\varepsilon^2} \frac{\sum_{a} e_a^2 
\tilde g^{a}_{1L}(x,z^2\btu^2) \tilde D^{a}_{1}(z,\btu^2)}{\sum_{a} e_a^2\tilde f^{a}_1(x,z^2\btu^2) \tilde D^{a}_{1}(z,\btu^2)}   
\, .\nn
\eea

Now we derive the formula to extract Bessel-weighted asymmetries by means
of an event by event weighting in $\qt$, while binning in $x$, $y$, and 
$z$.  
First we express  the unpolarized and doubly polarized helicity 
structure functions  in $\Bt$ space as
\bea
\mathcal{F}_{UU,T}&=&
\frac{1}{K(x,y)}\int d\Ph\Ph J_0(\btu\Ph)\left(\frac{d\sigma^+}{\dphi}
+\frac{d\sigma^-}{\dphi}\right)\nn\nn
\mathcal{F}_{LL}&=&\frac{1}{K(x,y)\sqrt{1-\varepsilon^2}}
\int d\Ph\Ph J_0(\btu\Ph)\left(\frac{d\sigma^+}{\dphi}
-\frac{d\sigma^-}{\dphi}\right)\, ,
\label{strucBspace}
\eea
using the shorthand notation for 
the differential phase space factor $d\Phi\equiv dx\, dy\, d\psi\,  dz\,  d\Ph P_{h\perp}$.  Re-expressing the cross sections in terms of the number of events
in the differential phase space ``volume'',  Eq.~(\ref{strucBspace}) is given by, 
\bea
\mathcal{F}_{UU,T}&=&\frac{1}{K(x,y)}
\int d\Ph\Ph J_0(\btu\Ph)\left(\frac{1}{\mathcal{N}_0^+}\frac{dn^+}{d\Phi}
+\frac{1}{\mathcal{N}_0^-}\frac{dn^-}{d\Phi}\right)
\label{strucu}
\eea
and 
\bea
\mathcal{F}_{LL}&=&\frac{1}{K(x,y)\sqrt{1-\varepsilon^2}}
\int d\Ph\Ph J_0(\btu\Ph)\left(\frac{1}{\mathcal{N}_0^+}\frac{dn^+}{d\Phi}
-\frac{1}{\mathcal{N}_0^-}\frac{dn^-}{d\Phi}\right)
\label{strucl}
\eea
where $dn^\pm$ are the number of events in a differential phase space
volume, $d\Phi$, and $\mathcal{N}_0^\pm$
is the standard normalization factor, that is the product of the 
number of beam and target particles with $\pm$ polarization per unit target area.  In the following we assume that the experiment has been set up such that $\mathcal{N}_0^+=\mathcal{N}_0^-$.

Now we discretize the momentum integration in  Eq.~(\ref{strucu}) and ~(\ref{strucl}) for a fixed phase space cell in $x,y,z$ 
such that  the corresponding differential 
$dx\, dy\, dz$ becomes the bin volume $\Delta x\Delta y\Delta z$.
 Eqs.~(\ref{strucu}) and (\ref{strucl}) thus become 
\bea
\mathcal{F}_{UU,T}&=&
x\sum_ae_a^2\tilde{f}_1(x,z^2\btu^2)\tilde{D}_1(z,\btu^2)
\nn
&=&
\frac{1}{2}\left\{
\frac{1}{\mathcal{N}_0^+}
\sum_{i\, \epsilon\, {\rm bin}[x,y,z]}\, 
\frac{J_0(\btu \qti)\Delta n^+_i}{K(x,y)}
+\frac{1}{\mathcal{N}_0^-}
\sum_{i\, \epsilon\, {\rm bin}[x,y,z]}\, 
\frac{J_0(\btu \qti)\Delta n^-_i}{K(x,y)}
\right\}\frac{1}{\Delta x\Delta y\Delta z}\, , 
\nn
\label{FUU}
\eea
and
\bea
\mathcal{F}_{LL}&=&
x\sum_ae_a^2\tilde{g}_1(x,\btu^2)\tilde{D}_1(z,\btu^2)
\nn
&=&
\frac{1}{2}\left\{
\frac{1}{\mathcal{N}_0^+}
\sum_{i\, \epsilon\, {\rm bin}[x,y,z]}\, 
\frac{J_0(\btu \qti)\Delta n^+_i}{K(x,y)\sqrt{1-\varepsilon^2}}
-\frac{1}{\mathcal{N}_0^-}
\sum_{i\, \epsilon\, {\rm bin}[x,y,z]}\, 
\frac{J_0(\btu \qti)\Delta n^-_i}{K(x,y)\sqrt{1-\varepsilon^2}}
\right\}\frac{1}{\Delta x \Delta y\Delta z}.\nn
\label{FLL}
\eea
where we sum over the  discrete momentum index $i$, and 
$\Delta n^\pm_i$ are the number of events for polarization
$\pm$ as a function of $P_{h\perp i}$. 

Substituting Eqs.~(\ref{FUU}) and (\ref{FLL}) into Eq.(\ref{cspma}), 
 the experimental procedure to 
calculate the Bessel weighted asymmetry, $A^{J_{0}(b_TP_{h\perp})}_{LL}(b_T)$, 
becomes, 
\bea
A^{J_{0}(b_T\Ph)}_{LL}(b_T) &=&\frac{\tsigma^+( b_T)-\tsigma^-( b_T)}{\tsigma^+( b_T)+\tsigma^-( b_T)}\nn
&=& 
\frac{\displaystyle\sum_{{j}}^{N^{+}}\,  J_0(\btu \qtpj) 
-
\sum_{{j}}^{N^{-}}\,  J_0(\btu \qtmj)}
{\displaystyle\sum_{{j}}^{N^{+}}\,  J_0(\btu \qtpj)
+\sum_{{j}}^{N^{-}}\,  J_0(\btu \qtmj)}\nn
&\equiv&\frac{\tS^+-\tS^-}{\tS^++\tS^-}
\label{spma}
\eea
where $j$ are indices for the sums on events and $N^{\pm}$ are the number of events,
for positive/negative products of lepton and nucleon helicities and at given $x$, $y$ and $z$,
and where $\tilde{S}^\pm$ indicate the sum over events for $\pm$ helicities.

\section{\label{errors} Error calculations}

In this section we derive a formula for the standard deviation of the experimentally measured asymmetry Eqns. (\ref{spma1},\ref{spma}).  
First, we need to address sums of the form 
\be
	\tilde{S} = \sum_{j=1}^{N} J_0( \btu P_{h \perp j} )\ .
\ee
The number of events, $N$, can be regarded as a realization of a discrete random variable $M$ with a Poisson distribution. Our best guess for its expectation value $\mathrm{E}[M]$ is $N$. The momenta $P_{h \perp 1}, P_{h \perp 2}, ...$ are samples independently drawn from an unknown, continuous distribution. Thus the Bessel weights $J_0( \btu P_{h \perp 1}), J_0( \btu P_{h \perp 2}), ...$ are realizations of independent, identically distributed random variables $W_1, W_2, ...$ . The entire sum $\tilde S$ is thus the realization of a random variable
\be
	Y \equiv \sum_{j=1}^{M} W_j
\ee
This has the form of a \textit{compound Poisson distribution}, see, e.g., Refs. \cite{adelson1966compound,Bohm:2013gla}. The variance of $Y$ is known to be
\be
	\mathrm{Var}[Y] = \mathrm{E}[M]\, \mathrm{E}[W^2] = \mathrm{E} \left[ \sum_{j=1}^M W_j^2 \right]
\ee
So an estimate of the variance can simply be obtained from the recorded events by computing
\be
\mathrm{Var}[Y] \approx (\Delta \tilde S)^2 \equiv \sum_{j=1}^{N} J_0( \btu P_{h \perp j} )^2
\ee
Now we use the fact that the events for helicity products $+$ and $-$ are independent.
Therefore, the asymmetry $A^{J_{0}(b_T\qt)}_{LL}(b_T)$ given by Eqns. (\ref{spma1},\ref{spma}) receives two independent contributions to its uncertainty, 
\be
	\Delta \tilde S^+ = \sqrt{ \sum_{j=1}^{N^+} J_0( \btu \qtpj )^2 }, \qquad
	\Delta \tilde S^- = \sqrt{ \sum_{j=1}^{N^-} J_0( \btu \qtmj )^2 }\quad.
\ee
We then apply regular error propagation to obtain the standard deviation 
\bea
\Delta A^{J_{0}(b_T\qt)}_{LL}(b_T) & = &  
\frac{2}{\left(\tilde S^+ + \tilde S^-\right)^2}\sqrt{ (\tilde S^-)^2 (\Delta \tilde S^+)^2 + (\tilde S^+)^2 (\Delta \tilde S^-)^2  } \nonumber \\
& = & \frac{1-\left(A^{J_{0}(b_T\qt)}_{LL}(b_T)\right)^2}{2}
\sqrt{\left(\frac{\Delta \tilde S^{+}}{\tilde S^{+}}\right)^2 + \left(\frac{\Delta \tilde S^{-}} {\tilde S^{-}}\right)^2}\, .
\eea

\section{\label{bootstrap} Bootstrap technique for weighted Poisson events}

Consider a fit parameter $o$ from a fit to the asymmetries $A^{J_{0}(b_T\qt)}_{LL}(b_T)$ extracted for an array of values $\btu$. The fit parameter $o$ is just an example of an observable that is calculated from a set of intermediate results with strongly correlated statistical fluctuations. To perform an error analysis for $o$, we may turn to resampling techniques. 

Imagine we could repeat the entire experiment $K$ times, where $K$ is a large number, say $1000$ or $10000$.  We could calculate the observable for all $K$ experiments, resulting in values $o^{(1)}, o^{(2)}, \ldots, o^{(K)}$ and then compute the sample variance according to 
\be
	(\Delta o)^2 = \frac{1}{K-1} \sum_{k=1}^K \left(\  (o^{(k)})^2 - {\bar o}^2  \right)\ ,
	\label{bsvar}
\ee
where $\bar o \equiv K^{-1}\sum_{k=1}^{K} o^{(k)}$. This straightforward procedure gives us an estimate $\Delta o$ for the error of $o$ in the original experiment.
Bootstrap resampling \cite{Efron1979,nla.cat-vn372369} is a trick that allows us to do just that without repeating the experiment in reality. New data sets are produced from the original data using a random process that in some sense mimicks an actual repetition of the experiment.  \par 
In standard bootstrap resampling, the resampled data sets are all of the same size as the original data set. For the problem at hand, however, the sample numbers $N^+$ and $N^-$ are realizations of random variables as well. An obvious adaption of the usual resampling strategy is thus to vary the size of the generated data sets randomly, leading to the following algorithm:  
\begin{itemize}
	\item For $k$ from $1$ to $K$
	\begin{itemize}
		\item For both helicity products $\pm$
		\begin{itemize}
			\item Choose a random integer $N^{\pm(k)}$ according to a Poisson distribution with expectation value $N^{\pm}$.
			\item Choose $N^{\pm(k)}$ random integers $j_1^{(k)},\ldots,j_{N^{\pm(k)}}^{(k)}$ from a uniform distribution in the range $1 \ldots N^{\pm}$ (random sampling with replacement).
			\item Calculate $\tilde{S}^{\pm(k)} = \sum_{l=1}^{N^{\pm(k)}} J_0( \btu P^{[\pm]}_{h \perp j^{(k)}_l} )$\ for any desired value of $\btu$.
		\end{itemize}
		\item Calculate the asymmetries $A^{J_{0}(b_T\qt),(k)}_{LL}(b_T) = ( \tilde{S}^{+(k)} - \tilde{S}^{-(k)} ) / ( \tilde{S}^{+(k)} + \tilde{S}^{-(k)} )$ for any desired value of $\btu$.
		\item Calculate all other observables $o^{(k)}$ of interest using the asymmetries obtained in the previous step for data set $k$ . (This may involve fits to the $\btu$-dependence of the asymmetries.)
	\end{itemize}
	\item Determine error estimates for all observables $o$ using Eq. (\ref{bsvar}) or similar statistical means.
\end{itemize}
By performing the error analysis ``empirically'' for each observable $o$ in the very last step, correlations among the observables are taken into account correctly automatically. \par
Consider again the case that $o$ is obtained from a fit to the $\btu$-dependence. Depending on the details of the fit procedure (e.g., weights or covariance matrix) the size of statistical fluctuations of $o$ may vary.  However, as long as the resampled data is statistically representative, the value $\Delta o$ obtained from the bootstrap method will provide a good estimate of the expected fluctuations, appropriate for the chosen fit procedure.

The algorithm above is mathematically equivalent to the bootstrap method proposed in Ref. \cite{Bohm:2013gla},
even though the algorithm described in that reference looks different. 
According to Ref. \cite{Bohm:2013gla}, the instructions in the innermost loop of the above algorithm would instead read 
		\begin{itemize}
			\item Choose $N^\pm$ random integers $n^{\pm(k)}_1, \ldots, n^{\pm(k)}_{N^{\pm}}$ from a Poisson distribution with expectation value 1. 
			\item Calculate $\tilde{S}^{\pm(k)} = \sum_{j=1}^{N^{\pm}} n^{\pm(k)}_j J_0( \btu P^{[\pm]}_{h \perp j} )$\ for any desired value of $\btu$.
		\end{itemize}
To proove equivalence with the algorithm further above, let $\Poi_\lambda(n) \equiv \lambda^n e^{-\lambda} / n!$ denote the probability to obtain $n$ from a Poisson distribution with expectation value $\lambda$.  The probability to obtain integers $n^{\pm(k)}_1, \ldots, n^{\pm(k)}_{N^{\pm}}$ in the latter algorithm is
\bea
	\Poi_1(n_1)\cdots\Poi_1(n_N) & = & \frac{e^{-N}}{n_1! \cdots n_N!} = \Poi_N(T) \frac{T!}{N^T n_1! \cdots n_N!} \nonumber \\
	& = & \Poi_N(T)\ T! \prod_{j=1}^{N} \frac{(1/N)^{n_j}}{ n_j!} \nonumber \\
	& = & \Poi_N(T)\ \Multi_{T,N}( n_1, \ldots, n_N )
\eea
where we have ommitted the superscripts ${}^\pm$ and $^{(k)}$ for better readability, and where $T = \sum_{j=1}^N n_j$.
$\Multi_{T,N}( n_1, \ldots, n_N )$ is the probability to obtain the vector $(n_1, \ldots, n_N)$  from a multinomial distribution with $T$ trials and equal probabilities $1/N$ for all $N$ categories. Now if we identify $T = N^{\pm(k)}$, it becomes evident that the scheme of Ref. \cite{Bohm:2013gla} is indeed equivalent to first determining the number of terms $N^{\pm(k)}$ in the sum from a Poisson distribution, and then drawing momenta $\qtpmj$ randomly (with replacement) from the experimentally determined data set.

\bibliography{alu}

\end{document}